\documentclass[aps]{revtex4-2}
\usepackage{amsmath,amssymb,bbold}
\usepackage{color,graphicx}
\usepackage{geometry}
\newgeometry{tmargin=2cm, bmargin=3cm, lmargin=1cm, rmargin=1cm}
\newcommand{\be}{\begin{equation}}
\newcommand{\ee}{\end{equation}}
\begin{document}
 \title{Killing versus  branching: Unexplored   facets of  diffusive  relaxation.}
 \author{Piotr Garbaczewski  and Mariusz \.{Z}aba}
 \affiliation{Institute of Physics, University of Opole, 45-052 Opole, Poland}
 \date{\today }
 \begin{abstract}
 We analyze the  relaxation dynamics of   Feynman-Kac path integral kernel functions in terms of branching diffusion processes with killing. This sheds new light on the admissible path-wise  description of  the relaxation to equilibrium  for   conditioned Brownian motions, and diffusion processes with absorbing boundaries, where  Feynman-Kac kernels appear as the building blocks of inferred transition probability density functions.
  \end{abstract}
 \maketitle

\section{Motivation.}

The present endeavour departs directly from the discussion of various aspects of the relaxation vs killing intertwine in the path-wise analysis of the Brownian motion in trapping enclosures, c.f. \cite{zaba,gar,mazzolo0}.  That actually   stems from the pseudo-Schr\"{o}dinger reformulation of the Fokker-Planck dynamics, \cite{risken,pavl}, and the validity of the Feynman (respectively Feynman-Kac) path integration route in the derivation of integral   kernels of  closely  related motion operators $\exp(tL^*)$ and $\exp(-tH)$, \cite{hunt,zaba1,monthus,glimm}, c.f. also \cite{mazzolo, mazzolo1,stef}.

Here $L^*$ stands for the Fokker-Planck generator, while $H$ for the associated  Schr\"{o}dinger-type  Hamiltonian, \cite{zaba,zaba1,pavl}. We point out that the integral kernels in question are transition probability densities $p(y,s,x,t)=  [\exp(tL^*)](y,x),  0\leq s<t $ of the diffusion process  and  (Euclidean)  propagators $k(y,s,x,t) =  [\exp(-tH)](y,x)$ of the generalised Schr\"{o}dinger equation,  where $H=-(1/2)\Delta + {\cal{V}}$, and the potential  function  ${\cal{V}}(x)$   may take negative values on bounded domains,  while being bounded from below.

We focus on    Markovian  diffusion processes driven by    conservative (gradient)   time-independent  drift fields.  Let us consider a  diffusion process   $X(t)$, associated with the stochastic differential equation of the Langevin-type, (here interpreted in terms of infinitesimal  time increments)
 \be
dX(t) = b(X(t)) dt + \sqrt{2\nu } dW(t),
\ee
where  $b(x)$  stands for a forward drift,   $\nu$  is  a  diffusion constant ($2\nu $ is  interpreted as the variance parameter), and  $W(t)$  is the normalised   Wiener noise   in $R$, defined by   expectation values $\left< W\right> =0$ and $\left< W(s)W(t)\right> = \delta(s-t)$.

From now on we rescale the diffusion coefficient to the value $\nu =1/2$, to conform with the notation of \cite{zaba,zaba1,monthus}.  Accordingly, if an initial probability density   function  $\rho_0(x)$ is given,   its  time evolution
$\rho_0(x)= \rho (x,0) \rightarrow  \rho (x,t)=  [\exp(tL^*)\rho_0](x)$  follows  the  Fokker-Planck equation:
\be
\partial _t \rho = {\frac{1}2} \Delta \rho - \nabla (b \rho ) = L^* \rho ,
\ee
where the  operator $L^*=  \nu \Delta  - \nabla (b \, \cdot )$ is a  Hermitian $L^2(R)$  adjoint  of the, traditionally favored by mathematicians,   diffusion generator $L= \nu \Delta + b \nabla $, \cite{pavl}.

 We  anticipate the existence of a transition probability density function  $p(y,s,x,t)$, $0\leq s<t\leq  T$, ($T\rightarrow \infty $ is admissible)  for the diffusion process (1), (2):   $\rho (x,t) = \int p(y,s,x,t) \rho (y,s) dy$.  We presume $p(y,s,x,t)$ to be a (possibly fundamental) solution of the F-P equation, with respect to
   variables $x$ and $t$, i.e.  $ \partial_t  p(y,s,x,t)  = L^*_{x}p(y,s,x,t)$.

\subsection{Relaxation regime.}

Given $\rho (x,t)$ solving Eq. (2).  Let us  introduce an osmotic  velocity  field  $u =  \ln \rho ^{1/2} $  and  the current velocity field  $v=b - u$, with $b= - \nabla \phi  $, where $\phi =\phi (x)$ is time-independent.  We can readily  rewrite the F-P equation as the  continuity equation $\partial _t \rho = - \nabla j$, where  $j= v\cdot \rho $  has   a standard interpretation of a probability current.

We assume that the diffusion process asymptotically relaxes to the stationary (invariant)  strictly positive  pdf,   $\rho (x,t) \to \rho _*(x)$ as  $t\rightarrow \infty $.  In the  stationary  regime we have $j\rightarrow j_*=0$ and thence   $v\rightarrow v_*=0$. Since   $b $  is time-independent, the drift field  potential (presumed to be confining) $\phi (x)$  becomes correlated with $\rho _*$:  $b=u_* = \nabla  \ln \rho _*^{1/2} =  - \nabla \phi $.  Accordingly,   a stationary solution  of the Fokker-Planck equation actually appears in the form (Gibbs-Boltzmann by provenance, \cite{zaba})  $ \rho _*(x) = (1/Z) \exp[ - U(x)]$, with  the normalization constant  $Z = \int_R \exp(-U)dx$, where  $U(x) = 2\phi (x) $.

 Following a standard procedure \cite{risken,pavl}, given a stationary density $\rho _*(x)$,   one can  transform the Fokker-Planck  dynamics   into an associated Hermitian (Schr\"{o}dinger-type) dynamical  problem in $L^2(R)$,  by means of a  factorisation:
\begin{equation}
\rho (x,t) = \Psi (x,t) \rho _*^{1/2}(x).
\end{equation}
Indeed, the Fokker-Planck evolution  of  $\rho (x,t)$   implies the validity  of  the   generalized diffusion (Schr\"{o}dinger-type)   equation
 \begin{equation}
 \partial _t\Psi=  {\frac{1}2} \Delta \Psi - {\cal{V}} \Psi  = - H \Psi ,
\end{equation}
for  $\Psi (x,t)=  [e^{(-tH)}\Psi ](x)$, with $\Psi (x.0)= \rho (x,0)/ \rho^{1/2}_*(x)$ .

  Note that  the   $\rho (x,t) \rightarrow \rho _*(x)$ as $t\rightarrow \infty $,  needs to be  paralleled by  $\Psi (x,t)  \rightarrow  \rho _*^{1/2}(x)$, hence $\Psi (x,t)$  itself exhibits  the  relaxation  behavior (its path-wise implementation is actually the  main focus of  the present paper).

 We demand  that  $ H   \rho _*^{1/2} = 0$, which implies that the admissible   functional form of the  potential function  ${\cal{V}}(x)$   derives  as a function  of
 $\rho _*^{1/2}(x)$, \cite{zaba}:
\begin{equation}
{\cal{V}}(x) = {\frac{1}2} \,  {\frac{\Delta \rho _*^{1/2}}{\rho _*^{1/2}}} =     \frac{1}{2} \left(b^2  + \nabla b\right) = \frac{1}{2} \left[(\nabla \phi )^2  -  \Delta \phi \right] ,
\end{equation}
with  $b(x) = - \nabla \phi (x)$.  Note that proceeding in reverse, the functional form  (5) of the potential function  ${\cal{V}}(x)$ is a guarantee for  the  existence  of the bottom  eigenvalue  zero of the  Hermitian operator  $H= -{\frac{1}2} \Delta + {\cal{V}}$, associated  with a  strictly positive  ground state    $\rho _*^{1/2}(x)$.

We note, that by its very derivation, the potential function ${\cal{V}}(x)$ is not necessarily positive definite, nor non-negative,  but surely is bounded from below and continuous (this  is  secured by the properties of $\rho _*(x)$ and thence $\phi (x)= \nabla \ln \rho _*^{1/2}$).  This  admissible  negativity property  of ${\cal{V}}(x)$  on bounded subsets of $R$ will be of relevance in our further discussion. We shall relate it to the concept of trajectory cloning (branching) for killed diffusion processes.

\subsection{Path integration hints.}

Let us  notice that  by  employing  the  identity  $\nabla  (b \rho ) = (b  \nabla ) \rho +
\rho (\nabla b)$  we  can  rewrite the  F-P operator $L^*$, Eq. (2),  as follows
\be
L^* ={\frac{1}2} \Delta - b \nabla -  (\nabla b) =
{\frac{1}2}(\nabla  - b)^2  - {\cal{V}},
\ee
where   $\cal{V}$ has been previously defined in Eq. (5), c.f. \cite{zaba1}.

It is known, \cite{hunt,monthus,zaba1}, that the transition probability densities of the diffusion processes in question, actually coincides with  the  integral kernel  of the   motion operator   $\exp(tL^*)$:
\be
 p(\vec{y},s,\vec{x},t)= [e^{L^*(t-s)}](\vec{y},\vec{x}).
\ee
Moreover, \cite{hunt},   Fokker-Planck  transition probability density functions  and probability densities,
 for diffusions with (non)conservative drifts, are  known to be    amenable to   Feynman's  path integration
 routines. In case of  conservative drifts, this  can be achieved  by means of a   multiplicative (Doob-like)
 conditioning of the related (strictly  positive)   Feynman-Kac kernels,  \cite{zaba,gar,mazzolo0,hunt,olk,zaba1,glimm},
  provided the existence of   stationary  pdfs is granted.

The   path integral context  for  drifted diffusion processes   has been  revived in Refs. \cite{hunt,monthus,zaba1},
 through the formula   "for the propagator associated with the Langevin system" (1) (e.g.  the  integral kernel of the operator $\exp(tL^*)$:
\be
p(y,0,x,t)=  \exp(L^*t)(y,x)=  \int_{x(\tau =0)=y}^{x(\tau =t)=x}
   {\cal{D}}x(\tau ) \,  \exp \left[ - \int_0^t  d\tau {\cal{L}}(x(\tau ), \dot{x}(\tau )) \right],
\ee
where the $\tau $-dynamics stems  from  the (actually Euclidean)  Lagrangian ${\cal{L}}$:
\be
{\cal{L}}(x(\tau ), \dot{x}(\tau )) = {\frac{1}2} \left[ \dot{x}(\tau ) - b(x(\tau ))\right]^2  + {\frac{1}2}  \nabla b(x(\tau ))=
{\frac{1}2}\dot {x}^2(\tau )  - \dot{x}(\tau ) b(x(\tau )) + {\cal{V}}(x(\tau )),
\ee
with ${\cal{V}}(x)$ given by  Eq. (5).\\

{\bf Remark  1:}  We recall  that the "normal"   (e.g. non-Euclidean)  classical Lagrangian would have the form $L = T - V$ with $T=  \dot{x}^2 /2$   and $V(\dot{x},x,t)= {\cal{V}}   - \dot{x} b$.  The diffusion-induced  Lagrangian (9) clearly  has the  Euclidean  form   ${\cal{L}}= {\cal{T}} + V$.\\

Let us consider the action functional  (e.g. minus exponent) in Eq. (8),  in association with the  drift field $b = - \nabla \phi = \nabla  \ln \rho _*^{1/2}$.   We readily infer that the term $\dot{x}(\tau )\,  b(x(\tau ))$ in the Lagrangian (9) contributes:
\be
\int_0^t \dot{x} [- \nabla \phi (x(\tau ))]  d\tau =
  - \int_0^t {\frac{d}{d\tau }} \phi (x(\tau )) d\tau  = \phi (x(0)) - \phi (x(t))
\ee
to the action functional.

Therefore, the related probability density function  (path integral kernel of $\exp(tL^*)$) can be rewritten in  the form:
\be
p(y,0,x,t)=  e^{\phi (y) - \phi (x)}\,  k(y,0,x,t)
\ee
where the new function $k(y,0,x,t)$ is no longer a transition probability density (does not integrate to one) but an integral kernel of another  motion operator (actually $\exp(-tH)$, c.f. Eq. (4)):
 \be
 k(y,0,x,t) =   \int_{x(\tau =0)=y}^{x(\tau =t)=x}
   {\cal{D}} x(\tau ) \,  \exp \left[ - \int_0^t  d\tau {\cal{L}}_{st}(x(\tau ), \dot{x}(\tau )) \right],
\ee
where
\be
{\cal{L}}_{st} (\vec{x}(\tau ), \dot{\vec{x}}(\tau )) =
{\frac{1}2}\dot{\vec{x}}^2(\tau )  +  {\cal{V}}(\vec{x}(\tau ))
\ee
and ${\cal{V}}$ is given by  Eq.(5).

On the  operator  level, the passage from the  transition kernel $p$ of  (8) to $k$ of (12), amounts to the similarity transformation,  \cite{pavl,monthus,zaba,zaba1}:
\be
H = e^{\phi } L^* e^{-\phi }=  - {\frac{1}2} \Delta  + {\cal{V}},
\ee
which in fact "stays  behind"   the transformation (3),   mapping   the Fokker-Plack equation into the generalised heat (Schr\"{o}dinger-type) equation.
The outcome can be readily verified by resorting to the operator identity  $e^{\phi } \vec{\nabla } e^{- \phi } =  \vec{\nabla } - (\vec{\nabla }\phi )$.

Accordingly, we have  $[\exp (-tH)](\vec{y},\vec{x}) = k(\vec{y},0,\vec{x},t)$, whose  path integral evaluation reduces to the  Feynman-Kac formula (12), \cite{olk,glimm,klauder}.

 Concerning the generalised diffusion equation  equation  (4),  we clearly have   $\Psi (x,t)=  [e^{(-tH)}\Psi ](x)$ with H, Eq. (14).  It is useful to mention that for an undisputable validity of the formalism,  we  need to impose   some  assumptions  upon the  potential function ${\cal{V}}$: to be  continuous and  bounded from below  function,   plus an implicit technical assumption that $H$    is not merely Hermitian, but a   selfadjoint  operator, \cite{klauder,glimm}.  Then, we know that $k(y,0,x,t)=k(x,0,y,t)$ is   positive    symmetric  integral kernel   of the semigroup operator $\exp(- t H)$, given   by the Feynman-Kac formula  with an explicit ${\cal{V}}$   entry, c.f. \cite{gar,mazzolo0,klauder,faris}.  We emphasize that ${\cal{V}}$ may take negative values, while being bounded from below.

\subsection{ What does the Feynman-Kac formula tell us about the  trajectories   fate/destiny ?}

While in the  path integral vein, w recall that   the  Feynman-Kac  formula can be redefined    as  a weighted integral over sample paths  of the  Wiener process (colloquially, the free  Brownian motion), with the  conditional   Wiener path measure  $\mu _{(y,0,x,t)}(\omega)$    being involved, \cite{glimm,faris,klauder} :
\be
  k(y,0,x,t) =  [\exp(- t H)](y,x)  =
       \int   \exp[-\int_0^t {\cal{V}}(\omega(\tau  )) d\tau ]\,   d\mu _{(y,0,x,t)}(\omega).
        \ee
 Here  paths  $\omega $ originate from $y$ at time $t=0$ and their destination is $x$ to be reached  at time $t>0$).
In passing we note that  in contrast to the kernel function $k(y,0,x,t)$, transition pdfs  $p(y,0,x,t)$ are not symmetric functions of $x$ and $y$.

We may here  try to  imagine  a pictorial view  of the    Brownian motion in potential energy  landscapes, as   set by  Feynman-Kac potential spatial profiles. The Wiener path measure in Eq. (15) refers to paths of the free (undisturbed) Brownian motion, and it is the exponential factor which represents,  \cite{klauder},  "the distortion of the distribution of free-particle paths, introduced by the potential".
  Thus, a  possible  path-wise  interpretations of the Feynman-Kac formula  can be given   in terms of a
  random mover in  a   potential   ${\cal{V}}(x)$,  which  acts  as a mechanism that reinforces  or penalizes the
     random mover tendency   to  reside or  go into specific regions  of  space.  A "responsibility" for a  weighted  redistribution of random paths in  a given time interval,  is here transferred from drift fields  of Eq. (1)   to  the spatial   variability  of    potentials  ${\cal{V}}(x)$  of Eq. (5), specifically   to their curvature and steepness.

There is however a  problem. For the validity of the above pictorial view,  we should  presume that once released from $y$ at $t=0$, a  bunch of continuous   sample  trajectories   should be in existence (survive) up to the terminal point $x$ at time $t$. There should be  no loss or gain of the "probability mass", like e.g.  changes in the  overall number of involved sample paths,  or  surplus/deficit   contributions from  a priori admissible   paths with  random starting and terminal times, \cite{ito,helms,helms1,nagasawa,nagasawa1}.

 Essentially, if one accepts  the {\it realistic  particle} propagation ansatz, then  in the quantum mechanical contexts (the pseudo-Schr\"{o}dinger equation (5) being tentatively included)  nonrelativistic  particle  paths in a field of a potential  should never be terminated, \cite{klauder}.  We point out that  while  taking the "random mover" concept seriously,  the   killing picture described above  might be  appropriate when one relates the average to a diffusion process in a medium capable of absorption,   like e.g.  the diffusion of neutrons in a nuclear reactor with active  moderator materials.

On the other hand,  we  may invoke the {\it killing} alternative,  favored in Ref. \cite{ito}-\cite{nagasawa1}.
Then,  the Feynman-Kac formula is interpreted as a weighted  average over the Wiener process with weight $\exp[-\int_0^t {\cal{V}}(\omega(\tau  )) d\tau ]$  for each  sample  path.
 Let us tentatively assume, \cite{klauder,ito}, that ${\cal{V}}(x) \geq 0$,  departing for a while   from  the generic property of  the  (a priori)  confining  potential   ${\cal{V}}(x)$  to be bounded from below  (that  in principle allows ${\cal{V}}(x)$  to take negative values in  bounded  subdomains in $R$).

 We  may picture  a set of Wiener paths in the  $(x,\tau )$-space  in terms of a  random  {\it killing}  mechanism
  inflicted by  ${\cal{V}}(x)$. Namely, we admit that a particle following one of sample paths gets killed
at a point $x$, in the time interval $\delta \tau $ (infinitesimally  $d\tau $)  with the probability
${\cal{V}}(x) \delta \tau $.

The  killed  path is  henceforth  removed from the ongoing   (surviving) ensemble of Wiener paths.
That modifies the statistics of paths-in-existence  to the extent, that at  the final  time $t$,
the Feynman-Kac average is taken exclusively with respect to paths, which survive the full  period $[0,t]$ to
complete their travel from $y$ to $x$.

  The factor $\exp[-\int_0^t {\cal{V}}(\omega(\tau  )) d\tau ]$ in Eq. (18), is a probability that
a particle (random mover) complets its path from $(y,0)$ to $(x,t)$, c.f. \cite{klauder}.

Denoting $W(\tau )$  the Wiener process, c.f. (1),  we can write  a  formal
 stochastic differential equation for the   diffusion   process  with the  killing rate ${\cal{V}}(x)\geq 0$:
\be
X(\tau + d\tau ) =\left\{\begin{tabular}{cc}
	$\emptyset $ &   with probability ${\cal{V}}(X(\tau )) d\tau $,  \\
	$X(\tau ) +dW(\tau ) $ & with probability $[1-  {\cal{V}}(X(\tau )) d\tau ] $,
\end{tabular}\right.
\ee
Since the operator $H= -(1/2)\Delta + {\cal{V}} $ is the generator of the diffusion
process with killing, we recognize Eq. (4), with   $\Psi (x,t)= [\exp(-Ht)\Psi ](x)$,  as the  appropriate  motion rule (generalised diffusion equation) following directly  from  (16), \cite{mazzolo}. Surely, $\Psi (x,t)$ does not conserve probability, and for ${\cal{V}}> 0 $  asymptotically approaches $0$.

We emphasize, that  the  ${\cal{V}}> 0 $ ansatz temporarily excludes from considerations bounded from below potentials, which are negative-valued on finite open subintervals in $R$.

 We shall abandon this restriction in below, thus enforcing a compensation of killing via cloning of trajectories, and ultimately by  introducing branching diffusion processes in the context set by Eqs. (4), (5), (15).  To this  end we shall accomplish the path-wise construction of such processes, in conjunction with simple  Hamiltonian model system, naturally associated with  (1)-(5)  and  subsequently with (8), (12),  (15).

For further analysis, we select exemplary potentials  ${\cal{V}}(x)$  for the Hamiltonian (14),  in the form (5)  deriving  from the Fokker-Plack drift, $b(x)= - \nabla \phi =  \nabla \ln \rho _*^{1/2}$ : \\

 (i)   the downward  shifted harmonic potential
\be
{\cal{V}}(x) = {\frac{1}2} (x^2-1) = V(x) - {\frac{1}2},
\ee
where $1/2$ is the bottom eigenvalue of the standard quantum  harmonic  oscillator with $V(x)= x^2/2$.
The Feynman-Kac potential ${\cal{V}}(x)$, which is negative  in $(-1,1)$, derives from $b(x)= -x$, related to
$\rho _*(x)= \pi ^{-1/2} \exp(-x^2)$.  \\

 (ii) the downward  shifted infinite well potential
\be
{\cal{V}}(x)=\left\{\begin{tabular}{cc}
	$-\frac{\pi^2}{8}$,  &  $x\in(-1,1)$,  \\
	$\infty$,  &   $x\in R \setminus (-1,1)$.
\end{tabular}\right.
\ee
This  constant potential is  negative in $(-1,1)$, and can be obtained by shifting down  (energy renormalization) the standard infinite well potential by the lowest eigenvalue  $\pi ^2/8$  of the  energy operator  with potential bottom set at $0$, c.f.  \cite{zaba}  (the "standard" potential has the form (18), but takes the value $0$ in $(-1,1)$).   The  emergent  ${\cal{V}}(x)= -\pi ^2/8, x\in (-1,1)$ derives from the Fokker-Planck drift for the process confined in the interval forever,  $b(x)= - (\pi /2) \tan(x\pi /2)$, with $\rho _*(x)= \cos^2(\pi x/2)$ see e.g. \cite{zaba}.

We note that our exemplary  Hamiltonians have the bottom eigenvalue $0$, with   $\rho _*^{1/2}$  as the corresponding (ground state)    eigenfunction, compare e.g. also \cite{stef}.

  Our further discussion   will refer to  a less restrictive  conceptual setting, where   the sample path  notion will not necessarily refer to {\it  realistic }  particle  trajectories, but to more abstract sample
  paths of a stochastic diffusion process  (it is uselul to remember that  the mathematical construct
  of  Wiener paths  may be safely termed  "unrealistic"). This will ultimately lead us to the "branching alternative", as an appropriate   diffusion   scenario   underlying the dynamics (15) in case of  Feynman-Kac  potentials  with finite  negativity domains.

\section{Diffusion process with killing and cloning (branching).}

\subsection{Killing can be tamed:  harmonic potential.}

Since we know,  \cite{zaba,gar,mazzolo0}  (see below),  exact analytic formulas for  Feynman-Kac  integral kernels with  quadratic  potentials, we  shall  consider two  options concerning the choice of the potential  in Eq. (15): (i)  nonnegative, harmonic one    ${\cal{V}}(x) \rightarrow V(x)= x^2/2$,  and  (ii) bounded from below  ${\cal{V}}(x)= (x^2-1)/2 = V(x) - 1/2$, which  is  negative  in $(-1,1)$ (e.g. the "harmonic potential with subtraction", \cite{faris}). We keep  $y=0$  as the initial $t=0$  starting point  for all trajectories.

 This entails a  visualization of the drastic  difference in the  asymptotic  $t \to \infty $  behavior of $k(0,0,x,t)$, while set  against $k_0(0,0,x,t)$,   which is of major interest in our subsequent discussion,  and  ultimately will  lead us  to the concept of {\it  branching}  diffusion processes in the context of the dynamics  inferred from  (15).

\begin{figure}[h]
\begin{center}
\centering
\includegraphics [width=0.4\columnwidth] {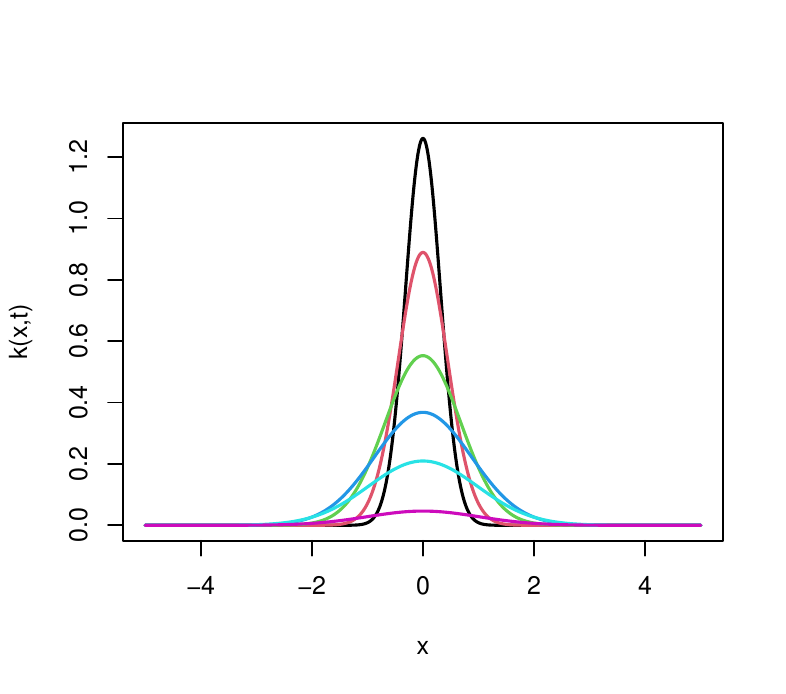}
\includegraphics [width=0.4\columnwidth] {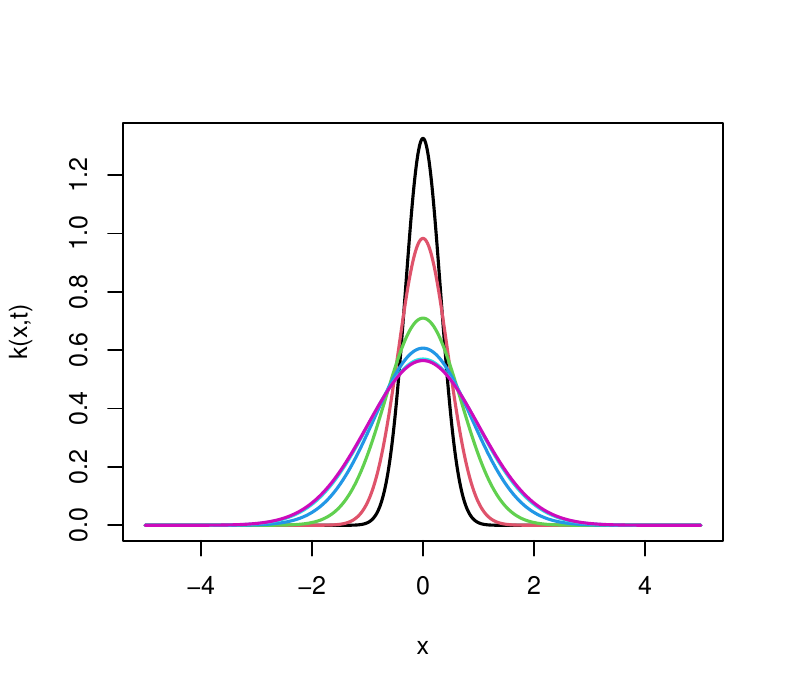}
\caption{Left panel:  Diffusion with the  harmonic $V(x)=x^2/2$ killing rate. An asymptotic decay down to $0$; follow the maxima of depicted curves $k(0,0,x,t)= k(x,t)$ for times $t=0.1, 0.2, 0.5, 1, 2, 5$.  Right panel: Relaxation process  with the Feynman-Kac potential ${\cal{V}}(x)= V(x)- 1/2= (x^2-1)/2$  for the same time instants as in the left panel.  The asymptotic curve is $K(x)= (\pi )^{-1/2} \exp(-x^2/2)$, with a maximum  ${\pi }^{-1/2} \sim 0,5641 $.}
\end{center}
\end{figure}

For clarity of discussion, we recall the analytic form of the integral kernel of the motion operator for the killed diffusion process:  $[\exp(-tH_0)](y,x)$ where $H_0= -(1/2) \Delta + V(x)$ and $V(x) = x^2/2$ is the standard harmonic potential.  We have,  \cite{zaba,gar,mazzolo0,glimm}:
\be
k_0(y,0,x,t) = \exp(- tH_0)(y,x)= {\frac{1}{(2\pi  \sinh t)^{1/2}}}  \exp  \left[ - {\frac{(x^2+y^2)\cosh t  - 2xy}{2\sinh t }} \right]  .
\ee

On the analytic level of description, the integral kernel of $[\exp(-tH)](y,x)$ with the renormalized harmonic potential  $V(x)=\frac{1}{2}(x^2-1)$  has the form looking  trivially different from the previous formula, since
the tamed killing effect is obtained via  a   multiplication  by $ \exp(+t/2)$.
\be
k(y,0,x,t)= \exp(- tH)(y,x)  = \exp(+t/2) k_0(y,0,x,t)=   [\pi (1-\exp(-2t))]^{-1/2} \exp \left[{\frac{1}2} (x^2-y^2) -   {\frac{(x- e^{- t}y)^2}{1- e^{-2 t}}}\right]  .
\ee

{\bf Remark 2:} We can here establish a direct relationship of the above propagators with the transition probability density $p(y,0,x,t)$ of the familiar Ornstein-Uhlenbeck process in $R$. Namely, we have
\be
p(y,0,x,t) =
k(y,0,x,t)\, {\frac{\rho_*^{1/2} (x)}{ \rho _*^{1/2} (y)}} = e^{+ t/2} k_0(y,0,x,t)\, {\frac{\rho_*^{1/2} (x)}{ \rho_*^{1/2} (y)}}=
(\pi [1-\exp(-2t)])^{-1/2} \exp \left[ -
   {\frac{(x- e^{- t}y)^2}{1- e^{-2 t}}}\right]
 \ee
where  the invariant pdf of the OU process reads  $ \rho _* (x)=  (\pi )^{-1/2} \exp(-x^2)$.    \\

Let us make a closer look at the  integral  kernel (20), which refers to the tamed killing.
By its definition,  at time $t=0$, all  sample   trajectories  are being  released from  $y=0$. Denoting $k(x,t)=k(0,0,x,t)$  we get
\be k(x,t)=\frac{1}{\sqrt{\pi(1-e^{-2t})}}\exp\left[\frac{x^2}{2}-\frac{x^2}{1-e^{-2t}}\right].
\ee
For large time values, and ultimately  $t\to\infty$,   the kernel  $k(x,t)$  asymptotically approaches
\be
K(x) =\frac{1}{\sqrt{\pi}}\exp\left(\frac{-x^2}{2}\right)
\ee
We note a subtle difference, namely   $(\pi )^{-1/2}$ replacing  $(\pi )^{-1/4}$,   if compared with the asymptotic form of $\rho _*^{1/2}$, recovered in the OU process.

Let us integrate the kernel (22)  over all locations  $x\in R$:
\be
K(t) = \int\limits_{-\infty}^\infty k(t,x)\,dx=\sqrt{\frac{2}{1+e^{-2t}}}.
\ee
Clearly $1\leq K(t)\leq \sqrt{2}$, and   as $t\to \infty $ , the upper bound  $K(t) \to K = \sqrt{2}\approx 1.41421$ is reached.  We note that $ K(t) $ is a monotonically  increasing function,  hence $\partial _t K(t) >0$.  To the contrary, we infer from (20) that $K_0(t)=e^{-t/2} K(t)$, decays exponentially for large $t$.  For completness, let us add that in the standard lore $K_0(t)\to 0$, with $K_0(0) =1$,  has the interpretation of the survival probabilityfor finite times $t$.

In below we shall give  computer-assisted  arguments to the meaning of  $K(t)$  as the quantitative measure of the  overall  number  $N(t)$ of  alive  sample trajectories, while set against their initial numer $N(0)$.  Actually, we shall demonstrate that $K(t) \approx N(t)/N(0)$.

We point out,  that this interpretation  is a straightforward  generalisation of the properties of $k_0(x,t)$ in case of  the  pure killing. We have verified that the  survival probability  $K_0(t) = \int _R k_0(x,t) dx$  provides a measure of the fraction of initially released trajectories, which have survived until tim $t$. Obviously, in the pure killing case, we have $N(t)/N(0) <1 $ for all $t>0$, followed by an asymptotic  decay to $0$

 Effectively,  as confirmed in simulations,   the initial  number  $100 000$ of released at $t=0$ trajectories,  in the course of the branching  diffusion with killing,  increases up to  $ \approx 141  421$.  This is  encoded  in the evolution of $K(t)$, which begins from $K(0)=1$ and approaches the limiting value  $K=1,4121=  141 421/100 000$.

  Thus, $K(t)$ is  the  relative  measure of the  "net  trajectory production surplus"  in our  branching  process with killing.

\subsection{A detour: killing and   cloning (branching) may saturate each other. }

We can legitimately  consider $K(t)$ of Eq. (24) as the $L(R)$ normalization of the  function $k(x,t)$, Eq. (22).
Accordingly, we get a legitimate probability density function
 \be
 \rho (x,t) = {\frac{k(x,t)}{K(t)}},
 \ee
 whose evolution rule directly follows  from the motion rule for $k(0,0,x,t)= [\exp(-tH)](0,x)$, (22), where $H= (1/2)[- \Delta +  (x^2-1)]$  and $\partial _t k= - H_x k$. We have:
 \be
 \partial_t \rho = {\frac{1}{K(t)}} \partial _t  k(x,t) -  k(x,t) {\frac{\partial _t K(t)}{K^2(t)}}  = - {\frac{1}2} \Delta \rho  - \left[ {\cal{V}}   +  {\cal{K}}\right] \rho
 \ee
 where we encounter a specific    time-dependent   killing-type  contribution ${\cal{K}}(t)$ to the overall expression for the potential term on the right-hand-side of Eq. (26), which is
 positive-valued for all $t>0$:
 \be
      {\cal{K}}(t) =  \partial _t\ln K(t) = + {\frac{1}{e^{2t} + 1}} .
 \ee
 We note that   $\rho (x,t)$ is a monotonically increasing  function  towards  the asymptotic shape  $K(x)/K$.  The "probability mass" $\int_R \rho (x,t) dx$   remains conserved and equals  $1$  for all times $t\geq 0$.

 If we interpret the positivity domain of  the original potential ${\cal{V}}(x) = (1/2)(x^2-1)$ as responsible for killing, while the negativity  domain as responsible for birth (cloning)  of trajectories, we  need to interpret  the negative-valued  "corrector"  ${\cal{K}}(t)$ in  Eq. (26)   as an additional  killing term, which reduces the surplus of cloned trajectories.  This secures that there is no "probability mass" excess,  we  have observed in connection with the large time asymptotic  of  $K(t)\rightarrow  \sqrt{2}$ ,  c.f.  Eq. (24).

 We remind that ${\cal{V}}(x)$ is negative in $(-1,1)$ and nonegative  in  $R \setminus (-1,1)$.  Thus the compensating term ${\cal{K}}$  increases the overall killing effect   on $R$  against that of cloning  alone  in $(-1,1)$.
As a consequence,  the random motion $\rho (x,t)$  not only preserves the "probability mass", but sets down at the asymptotic stationary  pdf
\be
\rho(x,t) \rightarrow   \rho _*(x)=  {\frac{K(x)}K} = {\frac{1}{\sqrt{2\pi }}} \exp \left(-{\frac{x^2}2} \right).
 \ee
which is a normalized Gaussian, with a maximum $(2\pi )^{-1/2} \sim 0.3989$.  This should be compared with the  $\pi ^{-1/2} \sim 0.5641$  outcome  mentioned in the caption of Fig.1.

We note that the limiting behavior (28) has close links with the concept of quasi-stationary distributions and  related  Yaglom-type limits,  \cite{huillet}-\cite{collet}.

\subsection{Direct   hint towards the branching scenario.}

In below we shall give a  computer-assisted  path-wise argument that  by integrating $k(x,t)$ and eventually $K(x)$ over $R$, we get a quantitative (relative)  measure of the number  of  alive  trajectories, if compared with their initial population at $t=0$.

Surely, the    integrals (23) and (24)  have   no meaning of the "survival" or "whatsoever" probability.  Nonetheless, (24)  admits a direct interpretation in terms of the ratio of the number of actually  alive trajectories, while set against   the initial number $100 000$ of released trajectories at time $t=0$.  In the harmonic case, an overall number of alive trajectories  increases up to a saturation limit, roughly about $141 421$.

This conjecture is   easily verifiable by means of our simulation  killing/branching/move-on  algorithm (detailed in below),  for all times of interest.  We can literally count all  trajectories crossing at $t$ any predefined  spatial subinterval $\Delta x$ in $R$.  For concretness, we  indicate that $\Delta x \approx  0,03$  is employed, while considering  the  interval of interest ($[-3,3]$ for reference), see e.g. Fig. 3.  In fact, the  coarse-graining  of the reference $[-3,3]$  needs a bit more meticulous approach, which we shall explain below in Section III, in the subsection devoted to the trajectory counting procedure.

 Accordingly,  the number of alive trajectories needs to grow in the process in which killing  (even if we regard it as tamed)  is admitted.  And that we can justify only by  introducing   the trajectory  birth  (cloning, branching) process. Ultimately, the  stable  balance between killing and branching ("probability mass generation") is achieved when we reach  (approximately) the "saturation"   number of  alive trajectories, (like $1.4 \cdot 10^5$, against the initial number    $10^5$ at  $t=0$).

 On the other hand, we can expand $K(t)$ defined in Eq.(24) into power  series for  small $t$,
\be
 \sqrt{\frac{2}{1+e^{-2t}}}=1+\frac{t}{2}-\frac{t^2}{8}+O(t^3).
\ee
The coefficient  $1/2$  in the linear term  may be interpreted as the  cloning/creation speed  for  new trajectories appearing in the branching eventssoon after the trajectory release form $y=0$ at $t=0$.  To justify this interpretation, we note that for  small $t$, all trajectories are  still concentrated in the close vicinity of  ($y=x(0)=0$), and they are effectively cloned  (alternatively, given birth) with the  probability
\be
-{\cal{V}}(x(t))\delta t=\frac{1}{2}(1-x^2(t))\delta t \approx \frac{\delta t}{2}.
\ee
We recall that  in the  interval $(-1,1)$ our potential ${\cal{V}}(x)= (x^2-1)/2$ takes negative values, whose sign inversion leads to the above probability notion.

For completness, let us mention that the potential (18) induces the appropriately modified version of the birth (branching) probability  (30):
\be
-{\cal{V}}(x(t))\delta t =  +\frac{\pi ^2}{8}  \delta t ,
\ee
which stands for the the birth (cloning, branching) probability for trajectories in the  interior $(-1,1)$ of the  interval  with absorbing ends. The above mentioned trajectory  counting procedure, in this case  involves a  coarse-graining of the interval $(-1,1)$.

\subsection{Interval with absorbing boundaries.}

Let us consider the interval $(-1,1)$ as a model arena for the diffusion process with absorbing boundaries at points $\pm 1$, c.f. \cite{zaba,gar,mazzolo0}, (we recall that  we assign the value $1/2$ to  the diffusion coefficient).
We reinterpret the original   diffusion  problem through a  useful  quantum-like artifice of the infinite well potential, and subsequently a related one,  whose well bottom is shifted  down on the energy scale  to  the  negative value    ${\cal{V}}(x)= - \pi^2/8$ for all $x\in (-1,1)$. We note that $+\pi ^2/8$  actually is  the ground state eigenvalue of the original quantum  infnite well spectral problem,  with the well bottom set at $0$.

The  Feynman-Kac  integral kernel  corresponding to the related (tamed killing)  dynamics has the form, \cite{zaba}:
\be
k(y,0,x,t)= \exp(\pi^2t/8)\, k_0(y,0,x,t).
\ee
and  like in the harmonic case,  differs from the standard killing kernel $k_0(y,0,x,t)$ by the killing  taming factor (here, $\exp(\pi^2t/8)$:
\be
 k_0(y,0,x,t)  = \sum_{n=1}^\infty \exp(-n^2\pi^2t/8)\,  \sin\frac{n \pi (x+1)}{2} \sin\frac{n\pi (y+1)}{2}
\ee

Since $\sin\frac{n\pi (y+1)}{2}$, for  $y=0$ equals $ \sin\frac{n \pi}{2}$, which  vanishes for $n$ even, the kernel $k(0,0,x,t)=k(x,t)$ can be rewritten as
\be
k(x,t)=\sum_{l=0}^\infty  (-1)^l \exp[(1-(2l+1)^2)\pi^2 t/8]  \, \sin \left(\frac{(2l+1)\pi (x+1)}{2}\right)
\ee
or equivalently
\be
k(x,t)=\sum_{l=0}^\infty   (-1)^l \exp[(-l^2-l)\pi^2 t/2] \,  \sin \left(\frac{(2l+1)\pi (x+1)}{2}\right).
\ee

For large times $k(x,t)$ approaches the asymptotic shape
\be
K(x)= \sin{\frac{\pi (x+1)}{2}} = \cos ({\frac{\pi x}{2}})
\ee
with $x\in (-1,1)$.

\begin{figure}[h]
\begin{center}
\centering
\includegraphics [width=0.4\columnwidth] {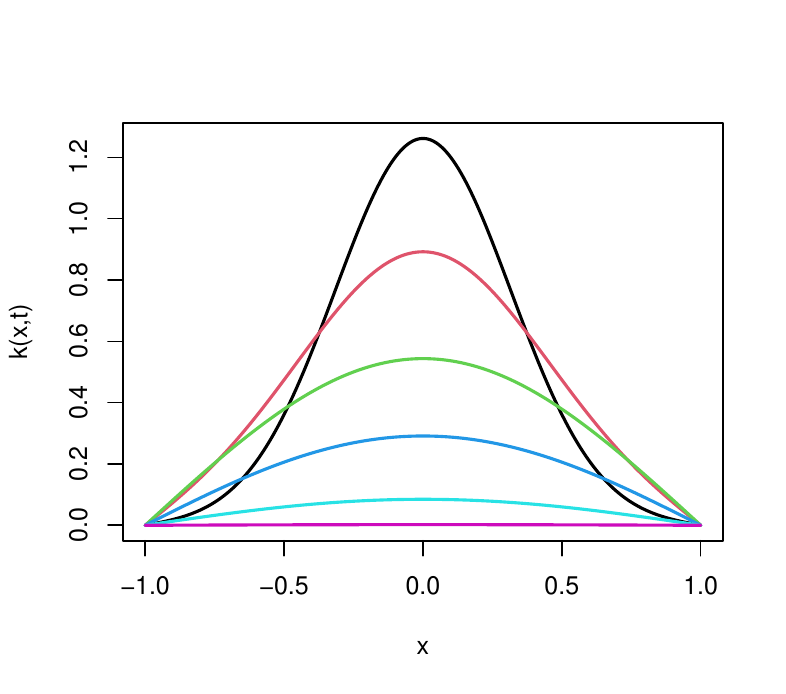}
\includegraphics [width=0.4\columnwidth] {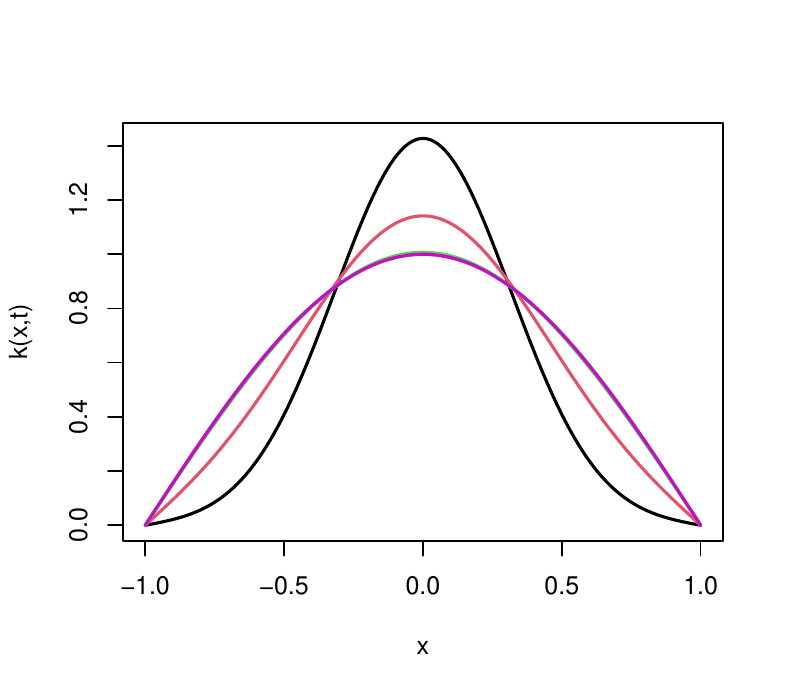}
\caption{Left panel: Diffusion   with killing (absorption) at the boundaries of the interval $(-1,1)$. An asymptotic decay  to $0$;  follow the maxima of depicted curves $k(0,0,x,t)=k(x,t)$ for times $t=0.1, 0.2, 0.5, 1, 2, 5$.
  Right panel:  Relaxation process  with the   Feynman-Kac potential (cloning  rate)  ${\cal{V}}(x)= + \pi ^2/8$ in $(-1,1)$,   for the same time instants as in the left panel.  The asymptotic curve is $K(x)=  \cos (\pi x/2), x\in (-1,1)$.  with a maximum  $1$.}
\end{center}
\end{figure}

Let us integrate $k(x,t)$ over $x$, and interpret the  outcome  as the quantitative  description  of   the a fate (destiny)  of a  bunch  (population)  of   random trajectories, while propagated from $t=0$ to some $t>0$.
 We have
\be
\int_{-1}^1 \sin \left(\frac{(2l+1)\pi (x+1)}{2}\right) dx = 4/[(2l+1)\pi],
\ee
and therefore
\be
K(t)=  \int_{-1}^1 k(x,t) dx = \sum_{l=0}^\infty \frac{(-1)^l 4}{(2l+1)\pi}\exp]-(l^2+l) \pi^2 t/2].
\ee
For large $t$ ($t\to\infty $) only the l=0  term survives  in $K(t)$ and  we have
\be
\lim\limits_{t\to \infty } K(t)  = K = {\frac{4}{\pi }}  \approx 1.2732 .
\ee

Clearly,  $\int _{-1}^{+1} \cos( \pi x/2) dx = 4/\pi $.    Therefore, we can proceed analogously to the discussion of Section II.B  and introduce the legitimate probability density function, see e.g. (25), with  the
 relaxation  behavior  manifested as  $t\to \infty $:
\be
\rho (x,t) = {\frac{k(x,t)}{K(t)}}  \longrightarrow   \rho _*(x)= {\frac{K(x)}K} = {\frac{\pi }{4 }} \cos( \pi x/2)
\ee
where $\rho _*(x)$ has a  maximum at  $ {\frac{\pi }{4 }} \approx 0.7854$.

If we turn back to the untamed killing case, i e.  consider the interval with absorbing ends,  a direct  conseqence of (32) and  (33)  is an exponential decay of the survival probability  $K_0(t)= e^{-\pi ^2/8} K(t) \approx  (4/ \pi ) e^{-\pi ^2/8} \rightarrow 0 $  with $t \to \infty $.  For the record, we  infer from Eqs. (32), (33) the   large time  behavior of   $k_0(x,t) \rightarrow   e^{-\pi ^2/8} \cos( \pi x/2) \rightarrow 0$, c.f. the left panel of Fig. 2.

In line with the conclusion of Section II.A, we  tentatively interpret the asympotic  $K$ value of  $K(t)$ as a quantitative measure of the overall amoount of killed and (re)born  paths, that ultimately  survive.  We encounter the same as before  signature of the trajectory number   increase, due to the excess of cloning (branching) against killing.  Would we have  begun  with the number $100 000$  of initially released  trajectories,   their  net  number   should increase  to  about  $127 323$ as $t\to \infty $.  In terms of  $K(t)$  that is encoded  in a monotonic growth of $K(t)$  from $K(0)=1$   to   $K = (4/ \pi)   \approx 1.2732 $.

The necessary condition for the  validity of the  path-wise  diffusive implementation of this result, is that (i) the  killing (at the boundaries)   is hereby (over)compensated   by  the trajectory cloning (birth, branching) within the interval $(-1,1)$ , (ii) for large times the killing and cloning rates  saturate  each other, so that the stable  "survival probability" profile $k(x,t) \to K(x)$  is reached.

A computer-assisted analysis, which confirms the validity  of the above  killing/branching trajectory interpretation will be given in below.

\section{Diffusion with killing and branching: Direct path-wise analysis.}

\subsection{Trajectory generation.}

The trajectory (sample path) picture stems from the  standard Brownian motion  $\{X(t),t\geqslant 0\}$  (e.g. the Wiener process)  as  introduced in Section I,  and  next  incorporating killing via the stochastic differential equation (16).  As yet, we have left aside the cloning (branching) scenario.

 To implement a computer-assisted  trajectory interpretation  of the killed diffusion process  with branching, as outlined in sections I and II,  we  need to pass from the lore of  continuous nowhere differentiable  trajectories,  to their  space and time   discretised  approximants.  The simulation  procedure,  enabling the trajectories counting,  is  based on  standard  assumptions.

   Let $t\in[0,T]$, we set   $\delta t=T/n$ for a predefined value of $n\in\mathcal{N}$.  The notation $\delta t$ is fairly informal, but presupposes that  any  finite  time interval   $\delta t$ of interest  can be made arbitrarily small (we thus bypass the usage of $dt$).   The Brownian walk is defined according to $x(t+\delta t)= x(t)+\sqrt{\delta t}\cdot u$, where  $u$   is the random variable sampled from the normal distribution  $N(0,1)$, $x(0)=0$.\\

Our aim is to  construct a specific version of the diffusion process  (with a link to the broad subject of  branching random walks and processes, \cite{huillet}-\cite{collet}), where in the course of time we may have allowed  random  killing accompanied  by random  cloning (branching event, giving birth to the, not yet killed trajectory,   offspring)  of sample  paths.
Our primary motivation stems form the Feynman-Kac formula, known to be  valid for confining potentials $V(x)$, which may not necessarily be   non-negative (the essential restriction is that $V(x)$ is bounded from below and continuous in the area of interest).

 Our construction involves the  random   cloning (branching) option  for all sample  trajectories
in existence, provided they   visit (any time) the  potential  ${\cal{V}}(x)$ negativity area  $(-1,1)$. If  the trajectory visits the complement of $(-1,1)$ in $R$, it may be killed (terminated) at random. Trajectories are never killed in $(-1,1)$.

 We discretize time, as mentioned before (while properly adjusting   $n\gg 1$ for different test runs). If the simulated random  trajectory  takes   the value  $x(t)=x$  for some  $t\in [0,T]$, its subsequent  "behavior" admits three  instances: killing, cloning, and   moving on, whose realization    in  each   simulation step  $[t, t+ \delta t)$ depends on  the  concrete value of the  potential  ${\cal{V}}((x(t))={\cal{V}}(x)$, where the sign of ${\cal{V}}(x)$ is of particular importance.

 We adopt  the following killing/cloning/move-on  scenarios:\\

\noindent
(1) If  ${\cal{V}}(x(t))\geqslant 0$  we  interpret  $p(t)=\min(1, \delta t\cdot {\cal{V}}(x(t))$  as the  probability of the killing event at $x(t)=x$.  Depending on the killing outcome  we admit two options  for the step $[t,t+\delta t)$:\\
\indent
  (a) the trajectory is killed  at $x(t)=x$ with the probability $p(t)$,  and  thence removed from the  trajectory statistics at the  time $t+\delta t$. (For the interval with absorbing endpoints, each trajectory  entering the complement of $(-1,1)$  is killed with the probability one.)\\
\indent
  (b) if the trajectory is not  killed, then it moves-on, by  following the evolution rule  $x(t+\delta t)=x(t)+\sqrt{\delta t}\cdot u$, where  $u$ a random variable sampled from the normal distribution  $N(0,1)$,  (the trajectory  survival  probability at time $t$ is given by $(1-p(t))$).\\

\noindent
(2) If at $x(t)=x$,   the potential  is negative-valued,   we  consider   $|{\cal{V}}(x)| = -{\cal{V}}(x(t))$ as the  probability defining factor,  while  setting   $q(t)=\min(1, -\delta t\cdot {\cal{V}}(x(t)))$.  There is no killing in $(-1,1)$, and   we  admit  two options:\\
\indent
(a) the  cloning (branching) event - the trajectory clones itself (produces an offspring)  at  $x(t)=x $  with the probability $q(t)$, subsequently  both the clone and  the parent  trajectory independently   move-on  from the branching point,   in accordance with the adopted universal  rule  $x(t+\delta t)=x(t)+\sqrt{\delta t}\cdot u$, up to time $t+\delta t$. At $t+\delta t$  we thus need to handle two trajectories instead of  one.\\
\indent
(b)  no offspring -   the trajectory follows the evolution  $x(t+\delta t)=x(t)+\sqrt{\delta t}\cdot u$, (that with the  probability  $(1-q(t))$).\\

We emphasize that killing and branching options are mutually exclusive  in our procedure. (In the branching literature, c.f. \cite{huillet}-\cite{burdzy}, one may meet trajectories in which killing and   branching  occur  at the same space-time point).  We are interested in the statistics of  all "alive"  trajectories at  each (coarse-grained) time instant of time $t \in [0,T]$.

\subsection{Trajectory counting.}

In Fig. 3, 4  and 5 we display the  outcomes of an explicit trajectory counting  for   the considered before  harmonic  variants of the Feyman-Kac potential  ${\cal{V}}(x)$.  In below   we describe  the   adopted  counting/display procedure:\\

\noindent
(1) We coarse-grain the  spatial axis ($x$-label) by dividing the  reference  interval of interest  into small segments of length $\Delta x$.\\
 \indent
     (a) the reference interval (say $[-3,3]$ in the harmoonic case, or $(-1,1)$ for killing at boundaries) is selected as follows.  At a given  time $t$,  for all alive trajectories we choose  a minimal value   $x_{min}$ of a trajectory location, and likewise a maximal value $x_{max}$.\\
\indent
     (b) the obtained interval  $[x_{min}, x_{max}]$ is divided into $100$ segments, with length
     $\Delta_{t}x= (x_{max} - x_{min})/100$. We point out thet $\Delta_{t}x$ is specific for each chosen time instant $t=0.1, 0.2, 0.5, 1, 2, 5$ , and varies from time to time,  c.f. Figs. 3 and 4.\\

\noindent
(2) Once a time instant is selected, for each consecutive  segment $\Delta x$ covering $[x_{min}, x_{max}]$, we count  the number  $n(\Delta x)$ of simulated trajectories, which    reach  the pertinent subinterval.\\

\noindent
(3) To obtain the relative measure of the trajectory number increase/decrease,  we evaluate $h(\Delta x) = n(\Delta x )/ (\Delta x \cdot 10^5)$, which  is a quantitative measure of the  fraction of counted  in  a segment  $\Delta x $  trajectories,   while set  against their  initial  number  $10^5$,   per length of the subinterval.  The  number $h(\Delta x)$ corresponds to the height  of   the respective   vertical bar in Figs. 3 and 4.\\

\noindent
(4) We note that $h(\Delta x) \cdot  \Delta x $,   is the relative number of alive trajectories in $\Delta x$ at time $t$, and summing up over all $h(\Delta x)$ covering $[x_{min}, x_{max}]$     gives $N(t)/N(0)$ of Fig. 3, with $N(0)= 10 ^5$.\\

\noindent
(5)  The envolope  in   each drawing of Fig. 3 is  given by an exact analytic expression for $k_0(x,t)$, c.f. (19) and (20), at indicated instants of time.  The envelopes in Fig.4 are given by the analytic expression for $k(x,t)$, (22).\\

\textbf{Remark 3:} In our simulations, we choose  the reference time $T=5$  and  $n=50000$, hence  $\delta t=0.0001$. The initial number of released  trajectories equals $100 000$.  All  simulated  trajectories are started at $t=0$ at the point $x(0)=0$. The  outlined above trajectory  counting recipe  allows to estimate that in  Figs. 3 and 4 ,  the   spatial  coarse-graining subinterval  $\Delta x$   size  varies between $0.01$ and $0.05$.\\

\subsection{Harmonic killing versus branching.}

In Fig. 3 we  visualise the pure killing case with the killing rate  ${\cal{V}}(x)= x^2/2$. We  follow the  trajectory generation  recipe of Section III.A.
\begin{figure}[h]
\begin{center}
\centering
 \includegraphics[width=0.28\columnwidth] {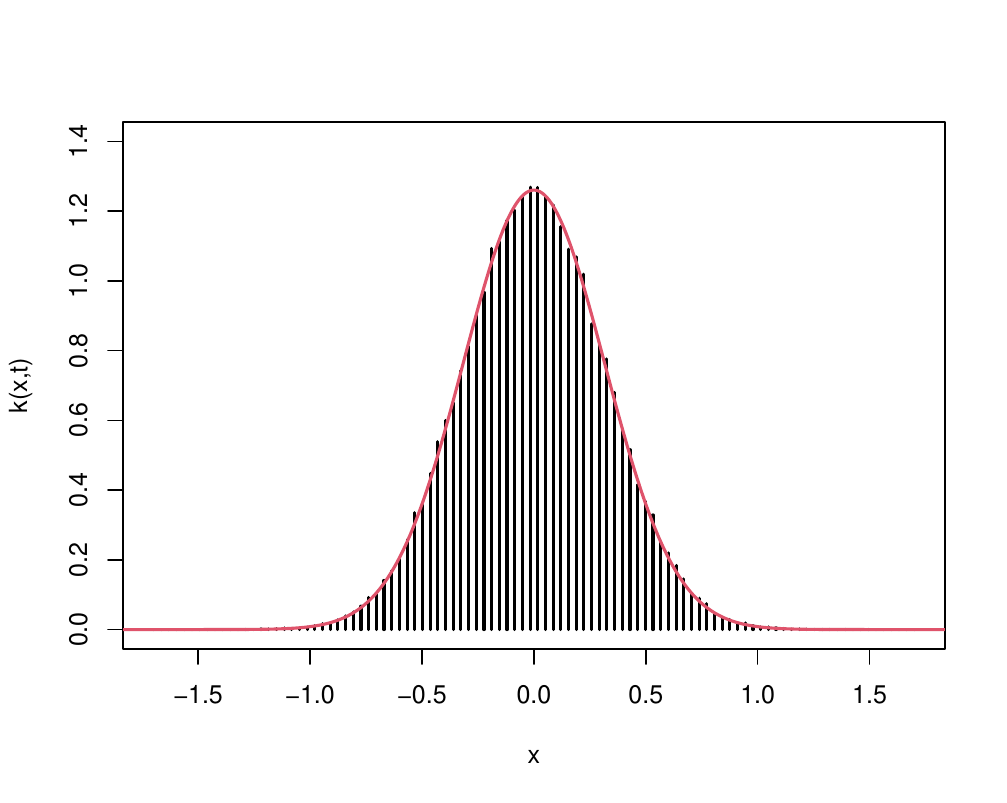}
\includegraphics[width=0.28\columnwidth] {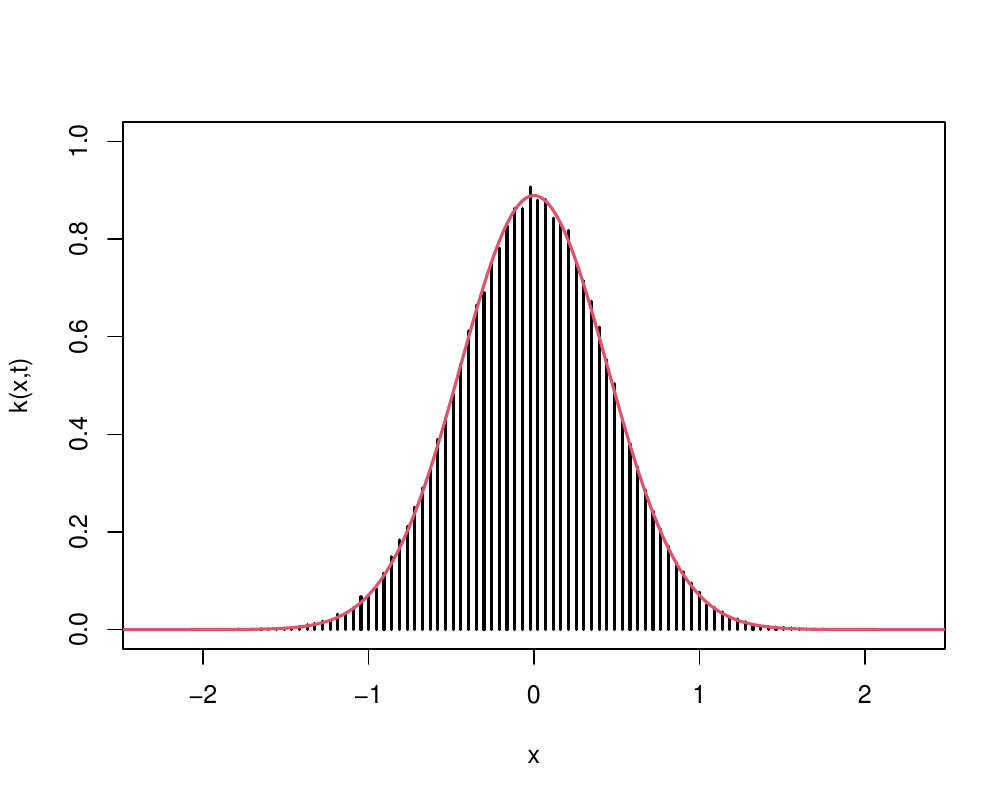}
\includegraphics[width=0.28\columnwidth]  {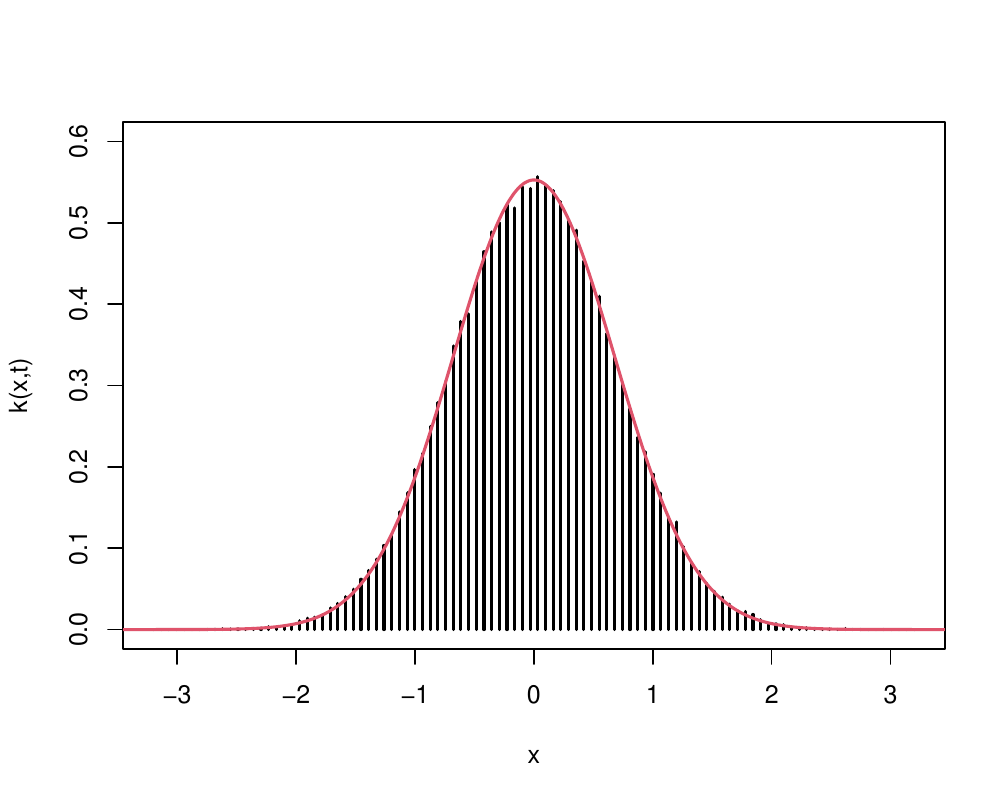}
\includegraphics[width=0.28\columnwidth]  {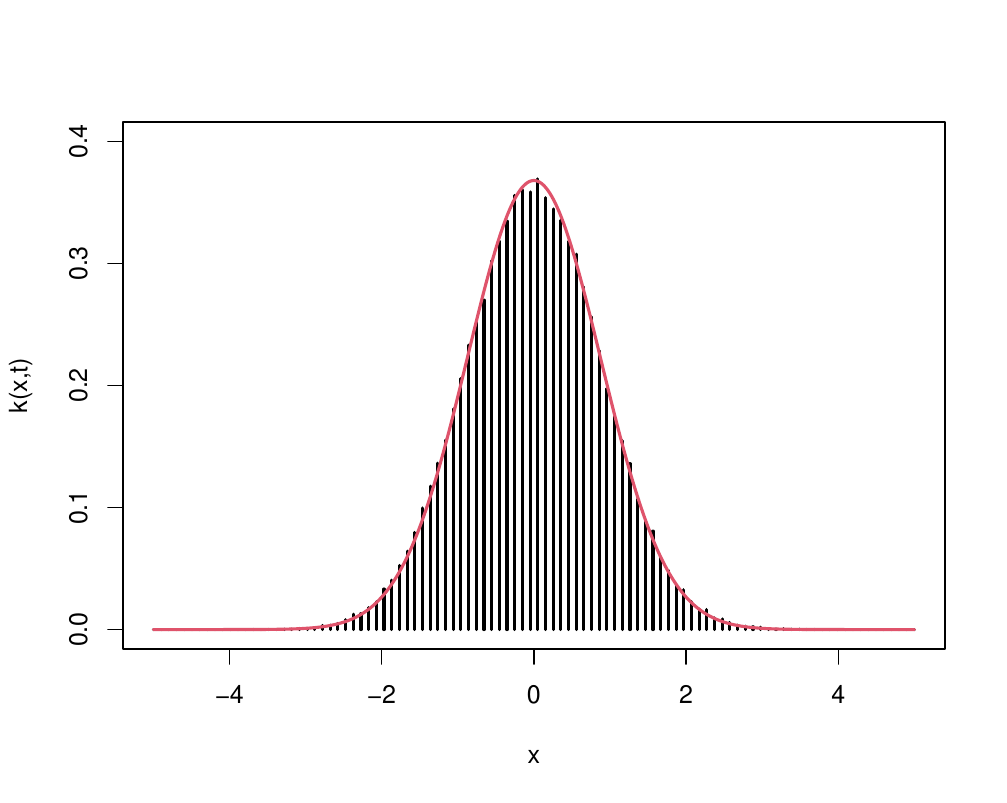}
\includegraphics[width=0.28\columnwidth]  {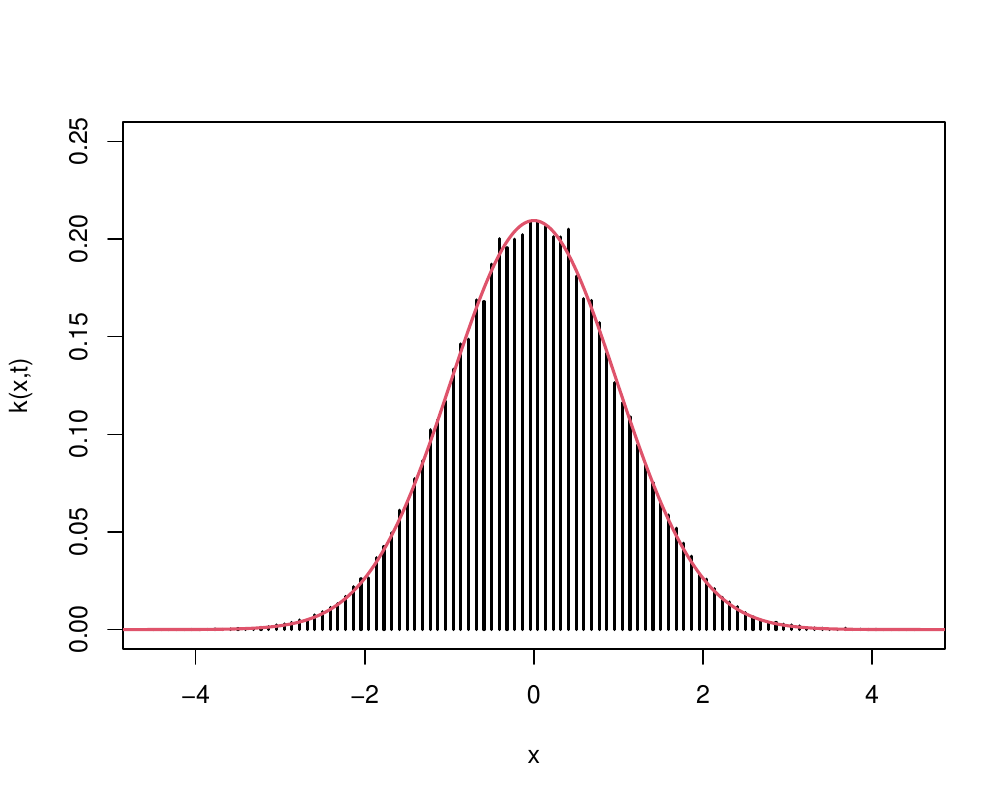}
\includegraphics[width=0.28\columnwidth]  {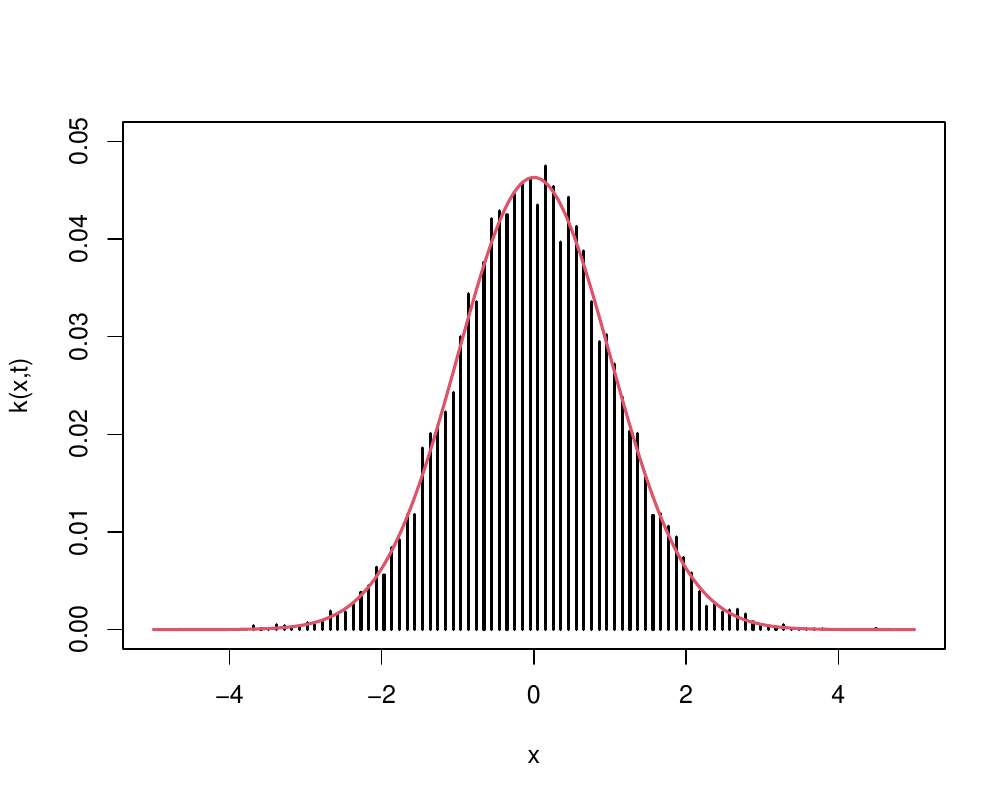}
\caption{Effects of the pure killing rate  ${\cal{V}}(x)= x^2/2$ in terms of the counting of surviving   trajectories  (we depict the recorded  numbers $N=N(t)$).  In both  rows, times in the increasing order from left to right. In the upper row: $t=0,1; N= 99766$, $t=0.2; N=98981$, $t=0.5; N= 94085$. In the lower row: $t=1; N=80462$, $t=2; N=51472$, $t=5; N= 11565$. Here  $N(0)= 10^ 5$. We realize that  $ K_0(t) = (\cosh t)^{-1/2} \approx N(t)/N(0)  \rightarrow 0$.}
\end{center}
\end{figure}
Note that scales  along the vertical axis change from panel to panel.  The  envelope (continuous curve) has an  exact  analytic form $k_0(x,t)= \exp(-t/2)\, k(x,t)$, with $k(x,t)$ given by Eq. (22).

The path-wise description of the interplay between killing (not allowed in $[-1,1]$) and branching (omnipresent in $(-1,1)$ is   comparatively  provided   in Figs.  4 and 5, by means of computer - assisted arguments (numerical simulation of trajectories, according to subsections III.A and III.B). We consider the Feynman-Kac potential ${\cal{V}}(x)= (x^2 -1)/2$.
\begin{figure}[h]
\begin{center}
\centering
 \includegraphics[width=0.28\columnwidth] {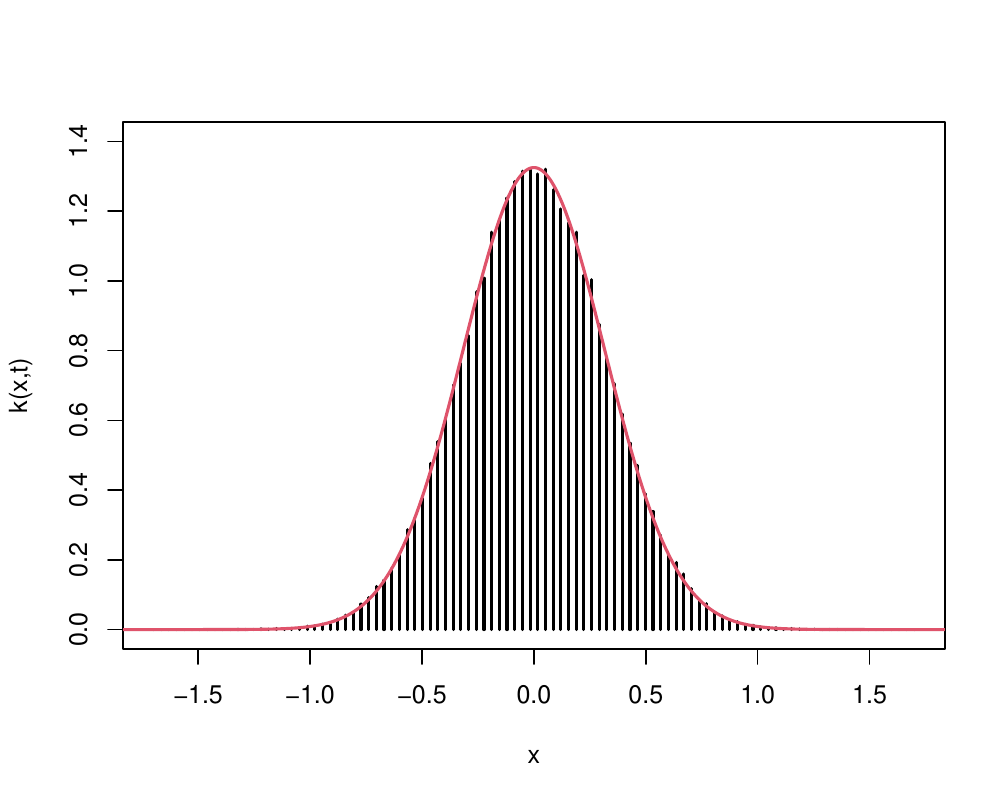}
\includegraphics[width=0.28\columnwidth] {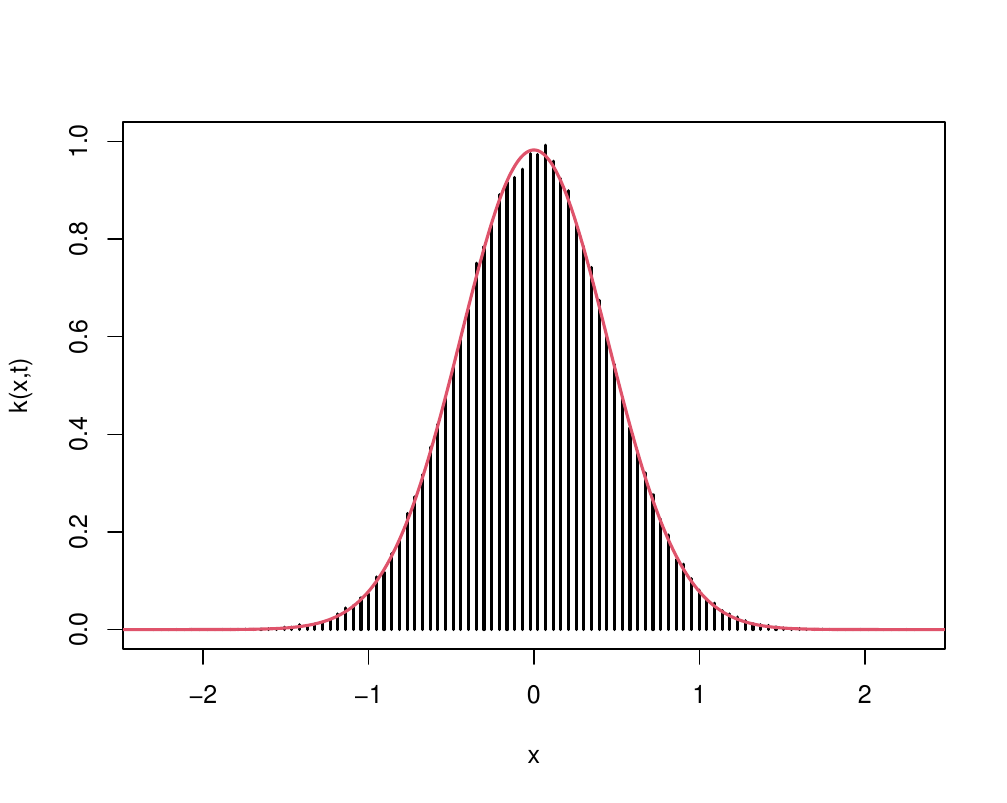}
\includegraphics[width=0.28\columnwidth] {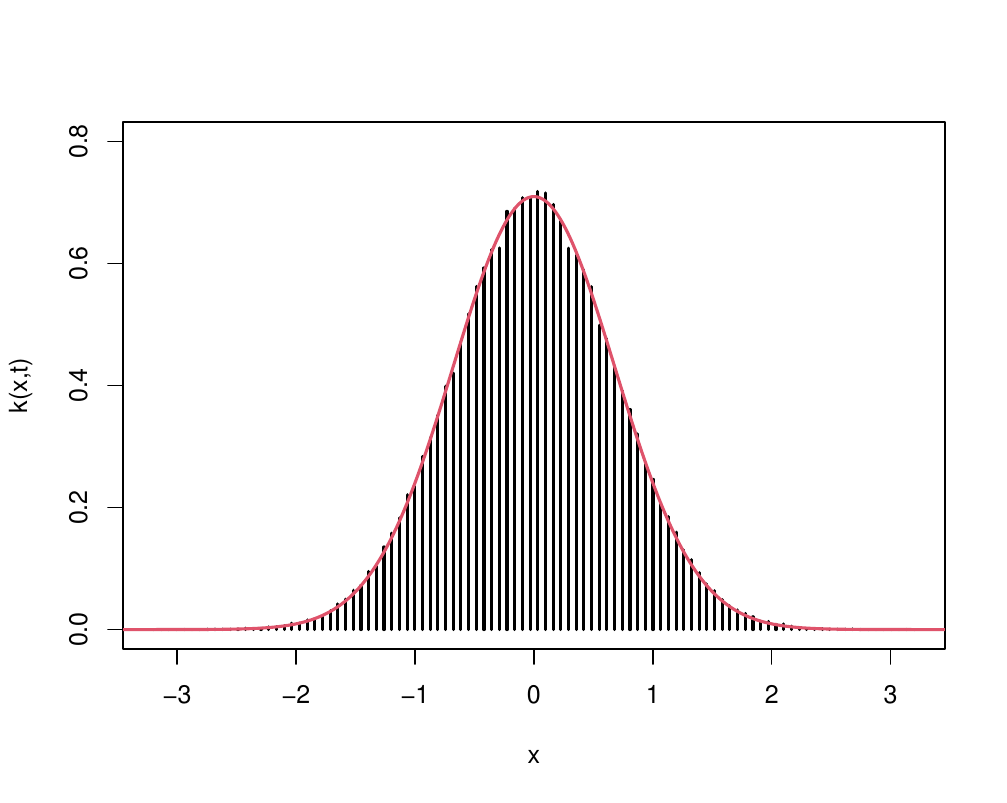}
\includegraphics[width=0.28\columnwidth] {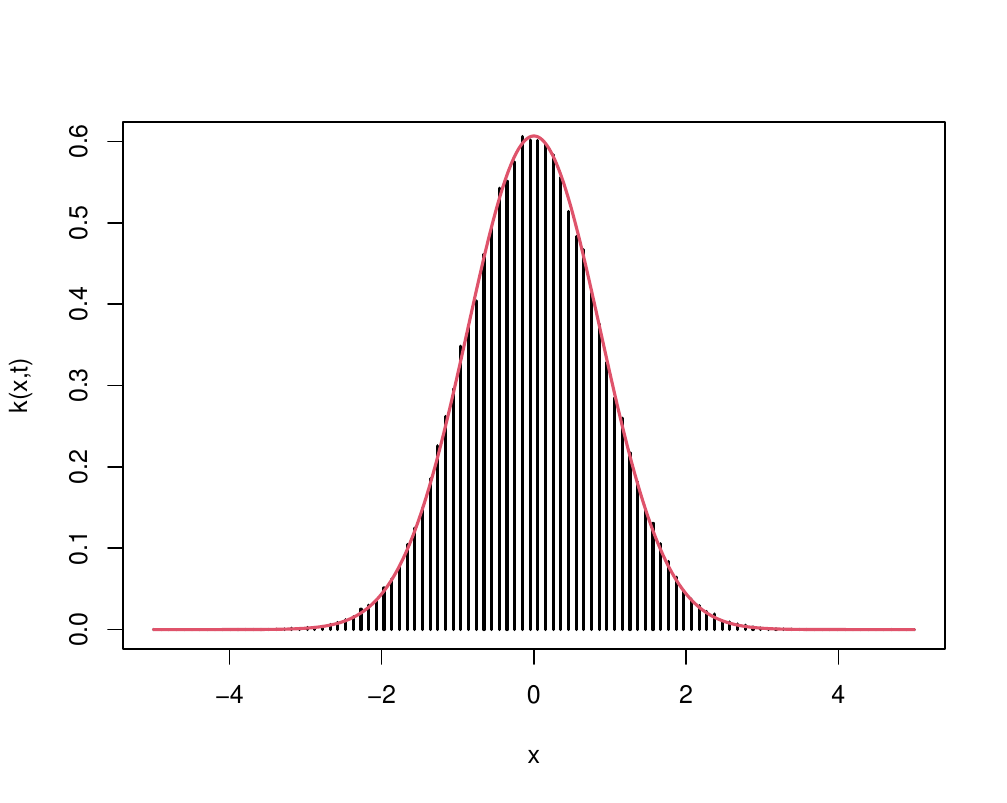}
\includegraphics[width=0.28\columnwidth] {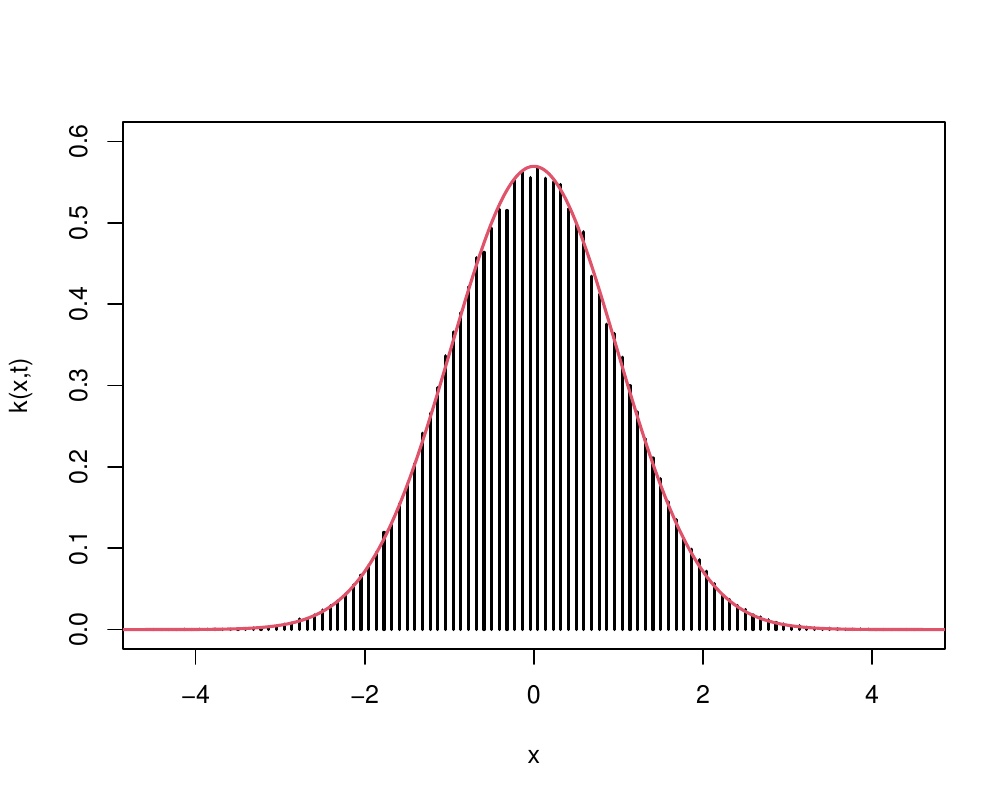}
\includegraphics[width=0.28\columnwidth] {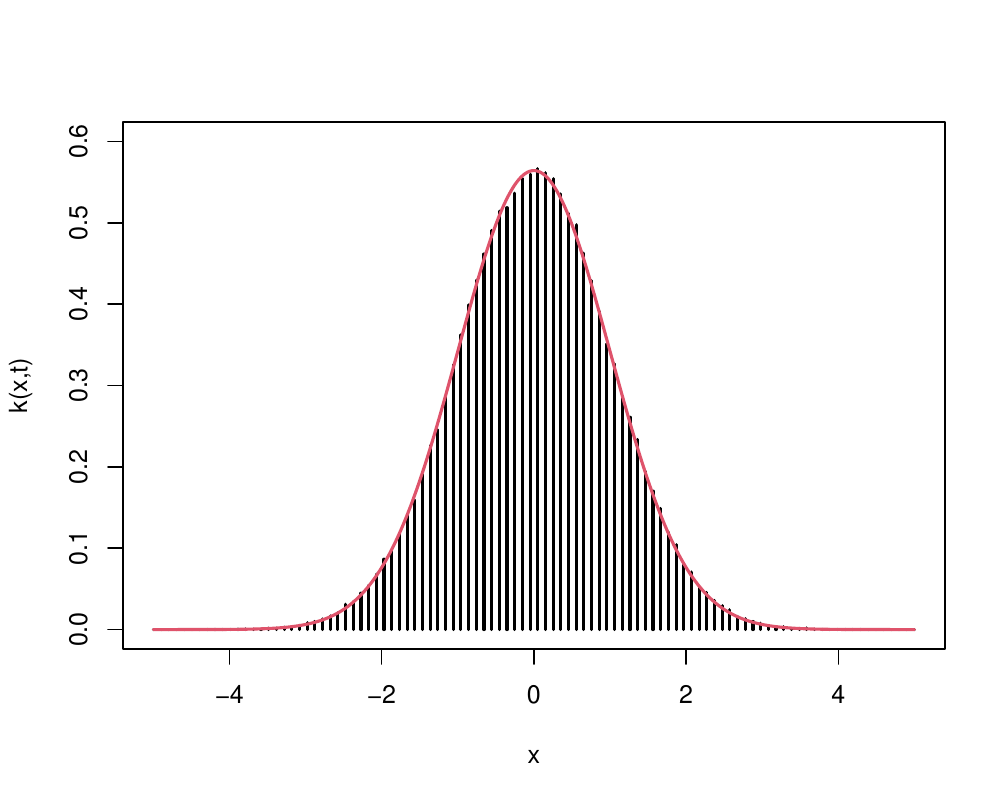}
\caption{Effects of the  (tamed)  killing rate  ${\cal{V}}(x)= (x^2 -1)/2$ in terms of   counted alive   trajectories  (we depict the recorded  numbers $N=N(t)$).  In both  rows, times in the increasing order from left to right. Note that scales  along the vertical axis change from panel  to panel.  The  envelope (continuous curve) has an  exact  analytic form $k(x,t)$,  as  given by Eq. (22).    Compare e.g. the right panel of Fig. 1.}
\end{center}
\end{figure}

\begin{figure}[h]
\begin{center}
\centering
\includegraphics [width=0.3 \columnwidth] {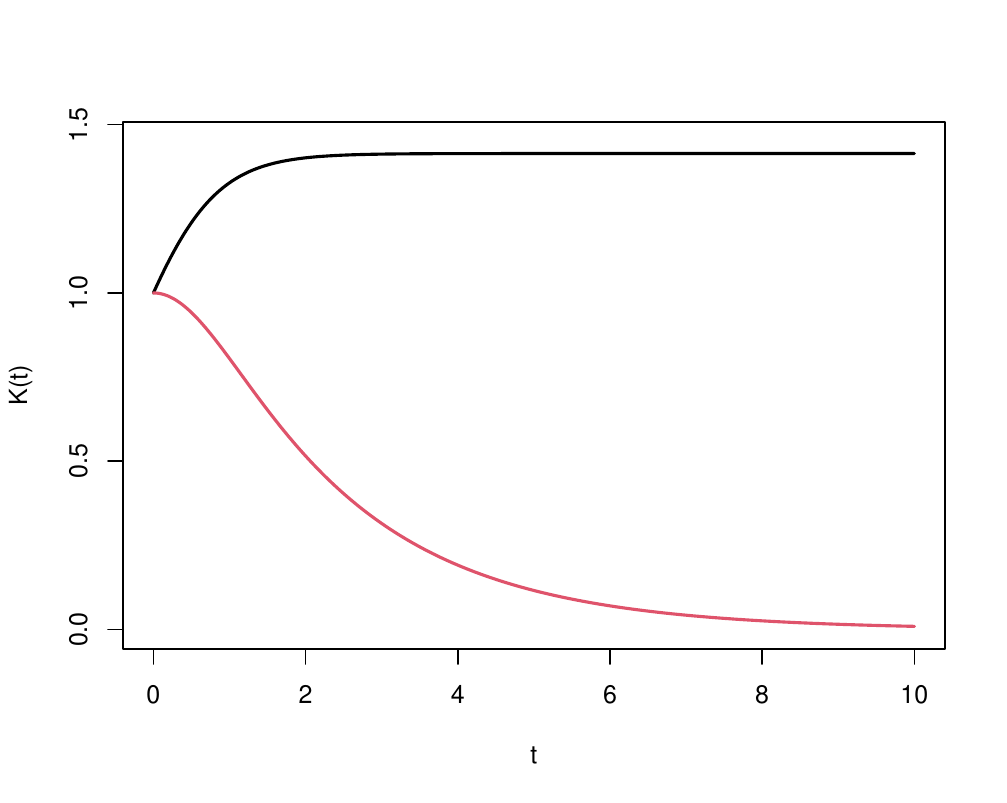}
\includegraphics [width=0.3 \columnwidth] {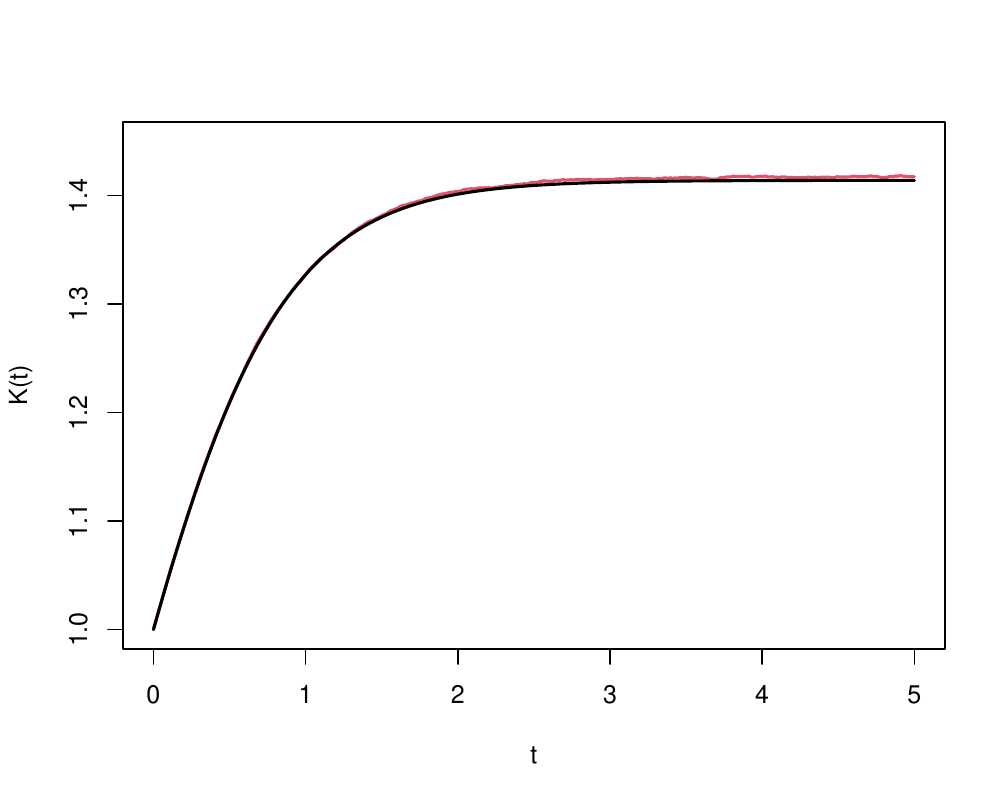}
\caption{Left panel:A comparative display of $K_0(t)$ (exponential decay) against $K(t)$ (relaxation).   We point out that $K_0(t)= \int _R k_0(x,t) dx$ integrates to $N(t)/N(0)  \approx K_0(t)= (\cosh t )^{-1/2}$,  c.f. Fig. 3.   Right panel: A comparison of the analytic formula for  $K(t)=  e^{(+t/2)} (\cosh t )^{-1/2}$  (black curve), while set against the numerically retrieved  curve  $N(t)/N(0)$  (red).}
\end{center}
\end{figure}
In connection with Fig. 5, we point out that   $K(t) \approx N(t)/N(0)$,   at the corresponding time instants.  In particular, $K(t)$ asymptotically approaches  $\sqrt{2}$, and accordingly   $ N(t) \rightarrow  10^5 \sqrt{2} \approx 141 421$.\\

\subsection{Killing versus branching for  the interval with absorbing ends.}

We begin from  the visualization  of the pure killing case, e.g. diffusion process in the interval $(-1,1)$  with absorbing ends  $\pm 1$.
In case of $t=5$,  the number of alive trajectories is too small to produce a reliable statistics (compare e.g.  $t=5$  harmonic  killing data in  Fig. 3.  Note that scales  along the vertical axis change from panel  to panel.  The  envelope (continuous curve) has an  exact  analytic form $k_0(x,t)$  given by Eq. (33).  For large times the envelope curve  scales down exponentially  to $0$, while preserving its functional shape:
  \be
  k_0(x,t)  \approx  e^{-t\pi ^2/8} \cos (x\pi /2)  \rightarrow 0,
  \ee
 c.f. also the left panel of Fig. 2.

\begin{figure}[h]
\begin{center}
\centering
 \includegraphics[width=0.28\columnwidth] {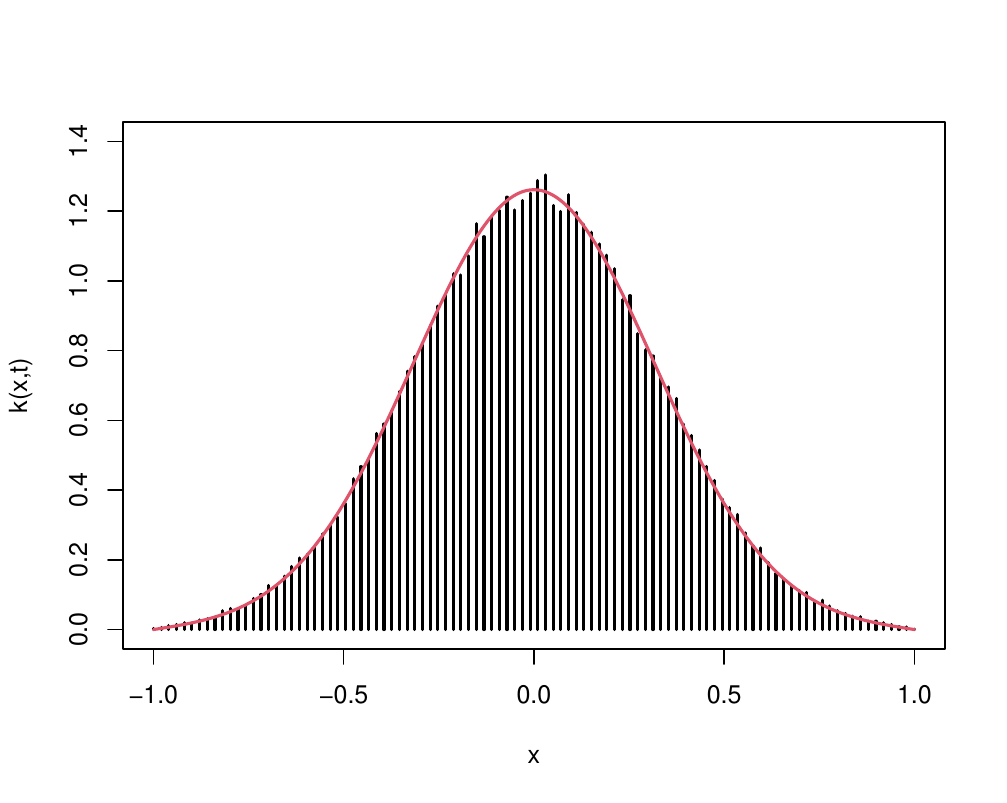}
\includegraphics[width=0.28\columnwidth] {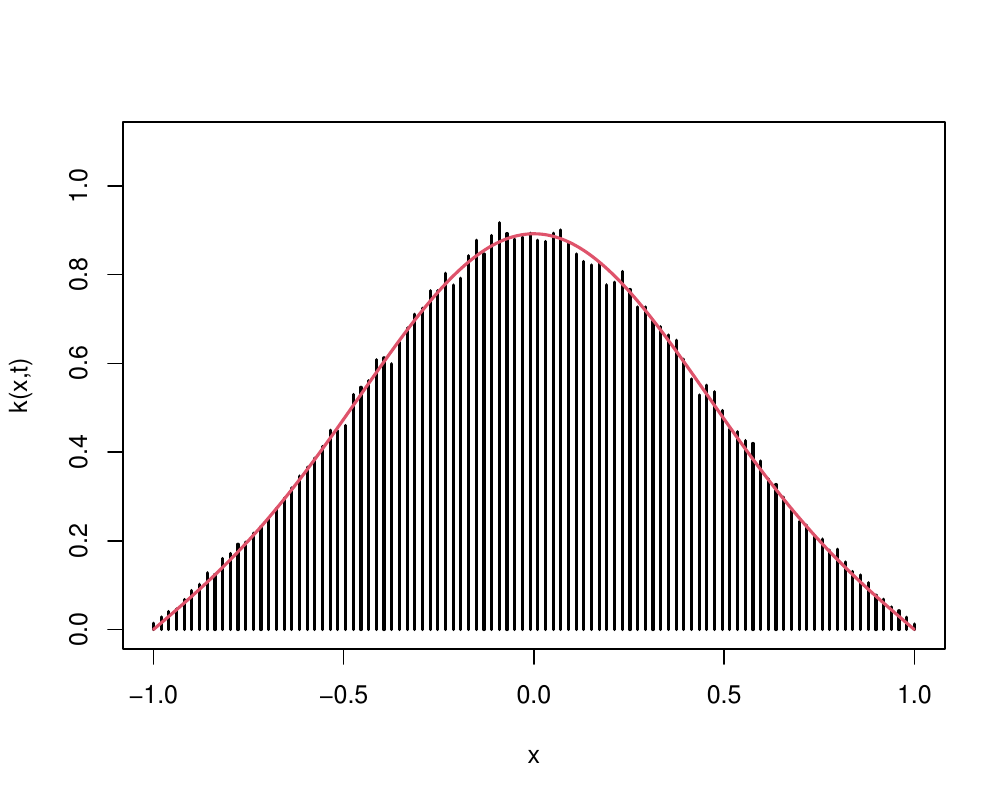}
\includegraphics[width=0.28\columnwidth]  {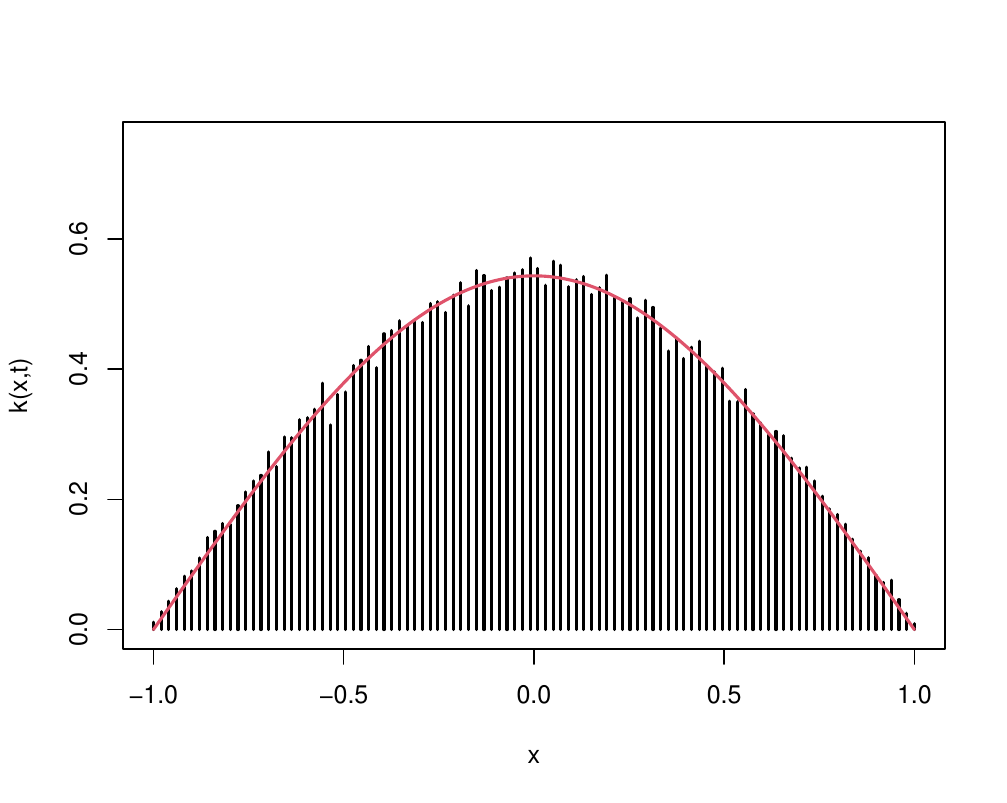}
\includegraphics[width=0.28\columnwidth]  {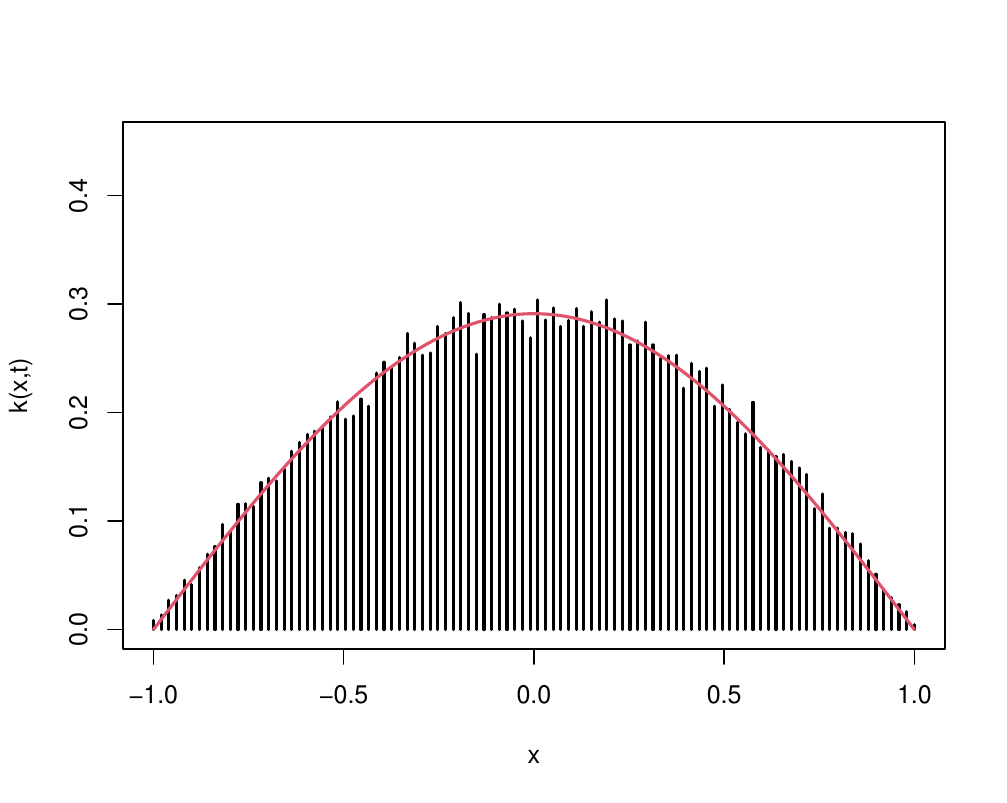}
\includegraphics[width=0.28\columnwidth]  {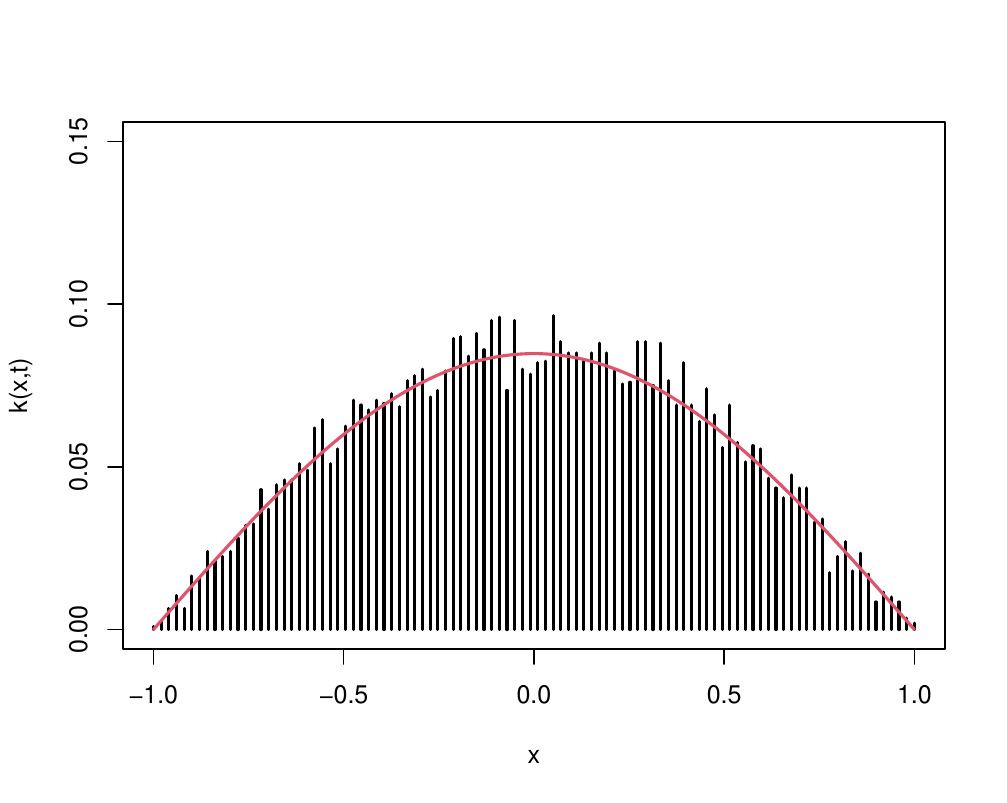}
\includegraphics[width=0.28\columnwidth]  {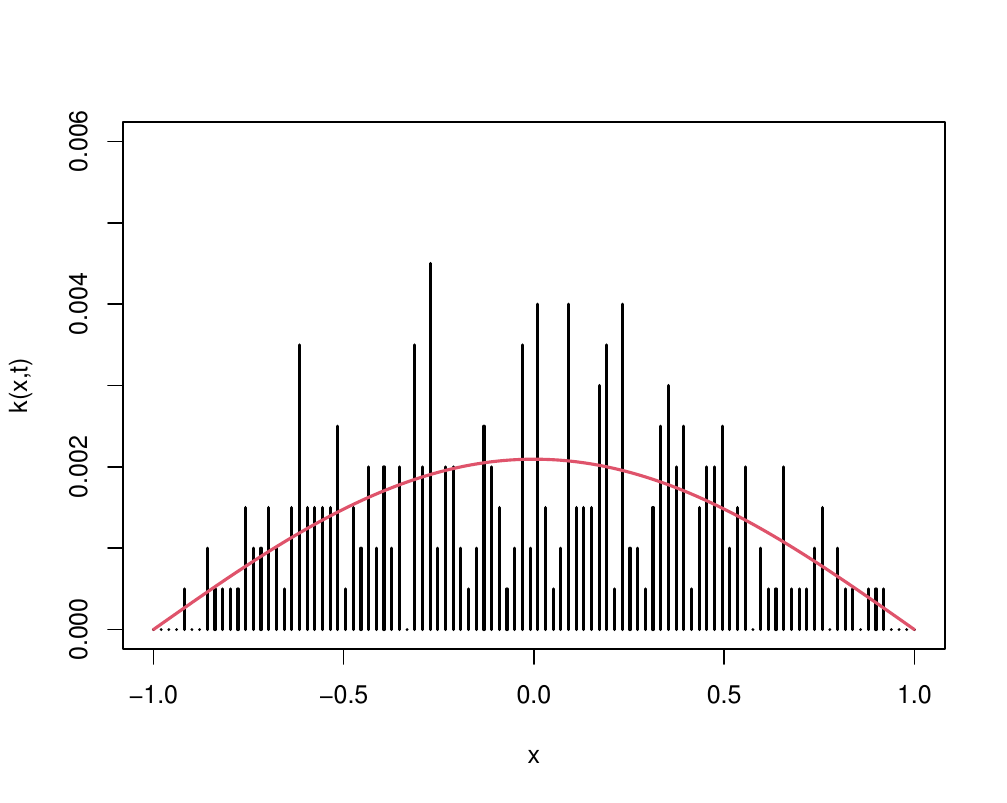}
\caption{Effects of the pure killing  for the interval with absorbing ends in terms of the counting of surviving   trajectories. We depict the recorded  numbers $N=N(t)$).   In both  rows, times in the increasing order from left to right. In the upper row: $t=0,1; N= 99 732$, $t=0.2; N=95 138$, $t=0.5; N= 69 086 $. In the lower row: $t=1; N= 37 535$, $t=2; N= 11 081 $, $t=5; N= 258$. We note that at $t=5$, the number of available trajectories is insufficient for the reliable statistics.}
\end{center}
\end{figure}

The   taming input of branching  upon killing   is visualized  for the branching rate choice  ${\cal{V}}(x)= - \pi^2/8$ in $(-1,1)$
\begin{figure}[h]
\begin{center}
\centering
 \includegraphics[width=0.28\columnwidth] {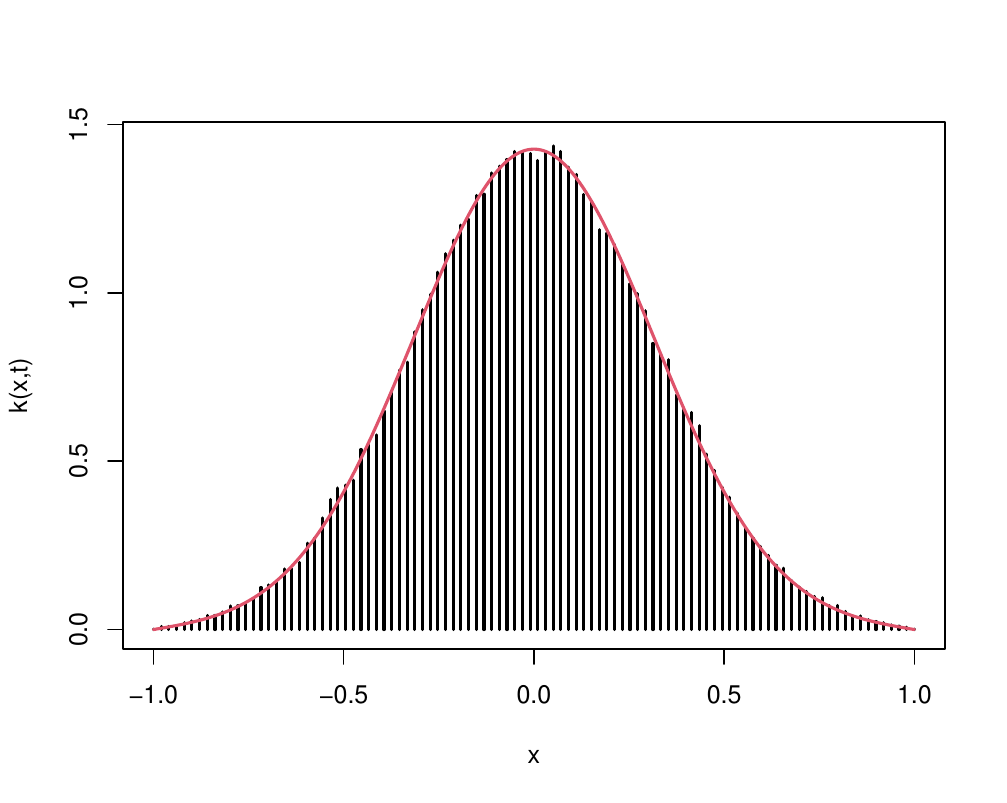}
\includegraphics[width=0.28\columnwidth] {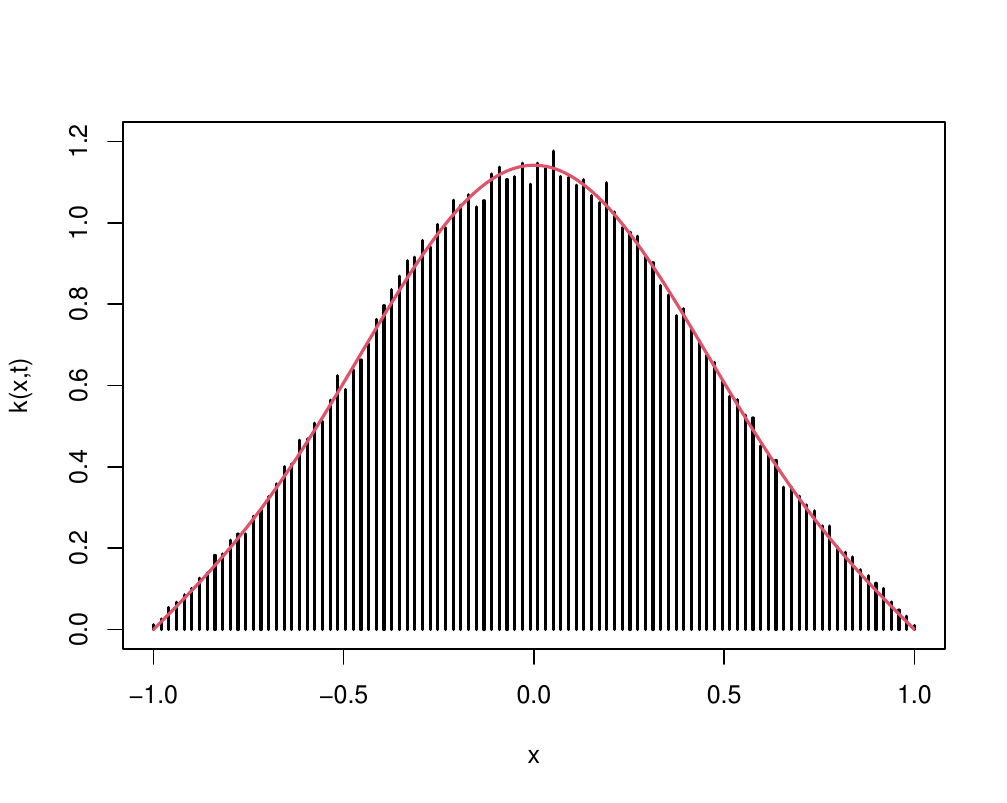}
\includegraphics[width=0.28\columnwidth] {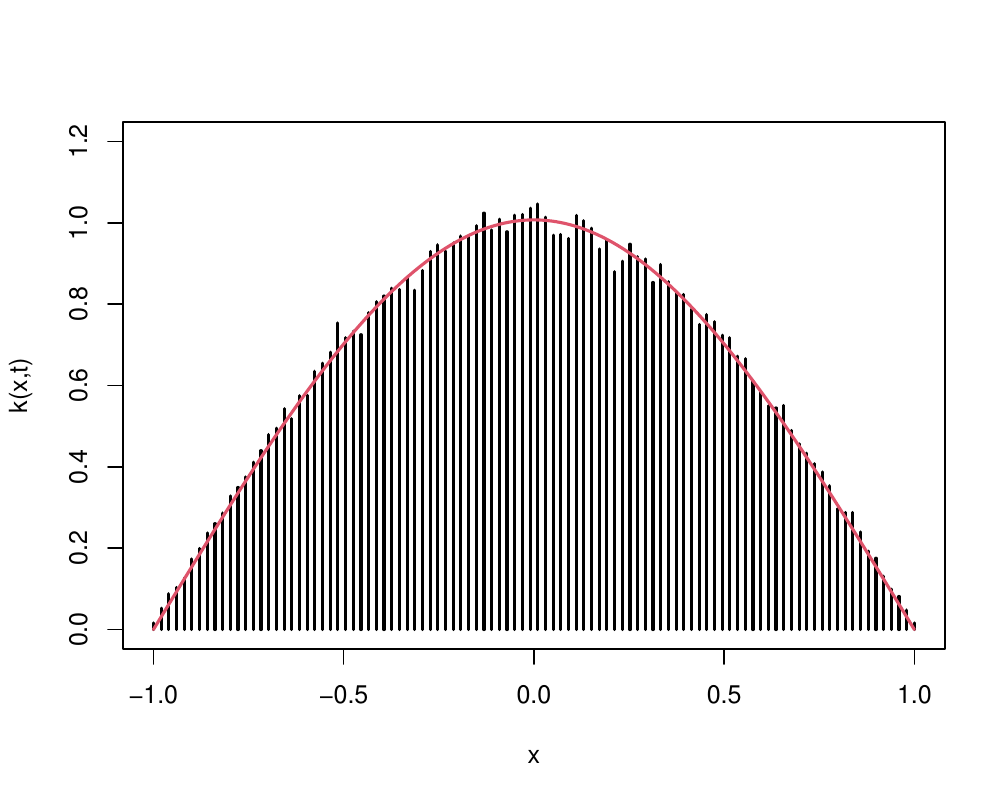}
\includegraphics[width=0.28\columnwidth] {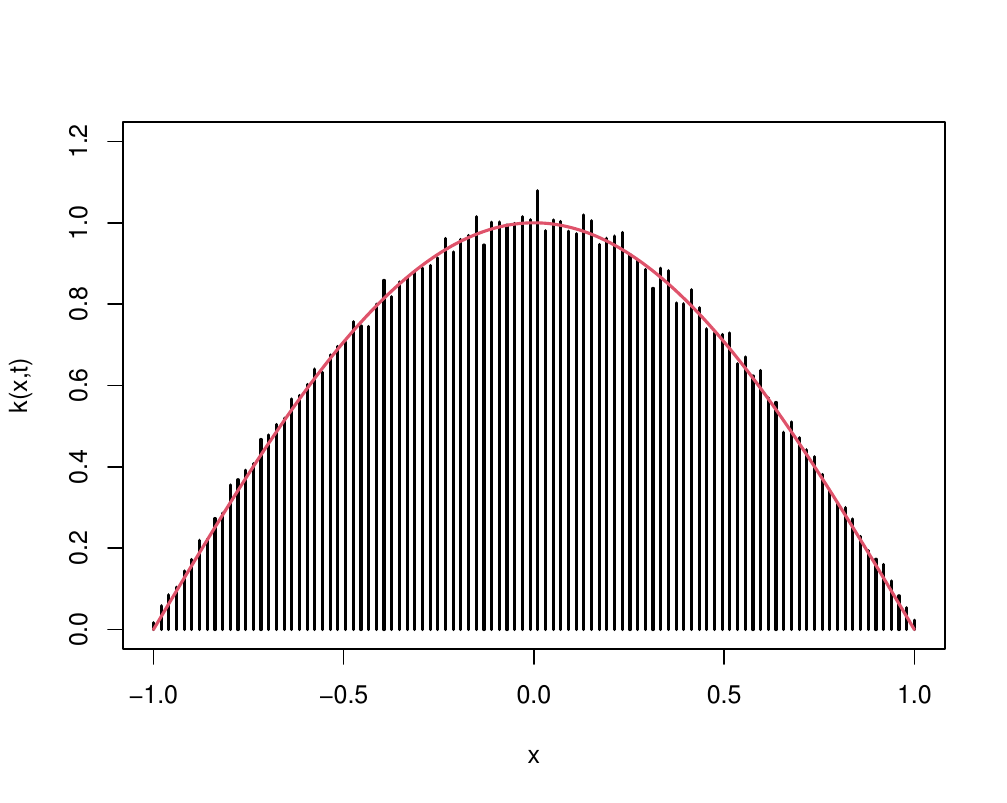}
\includegraphics[width=0.28\columnwidth] {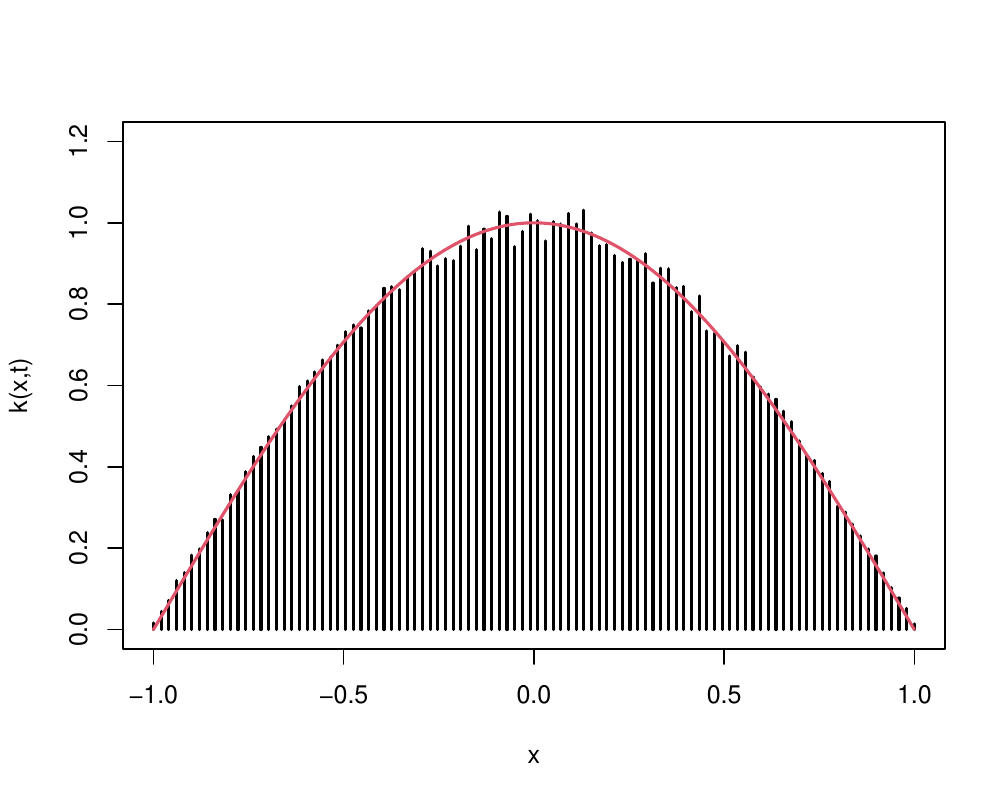}
\includegraphics[width=0.28\columnwidth] {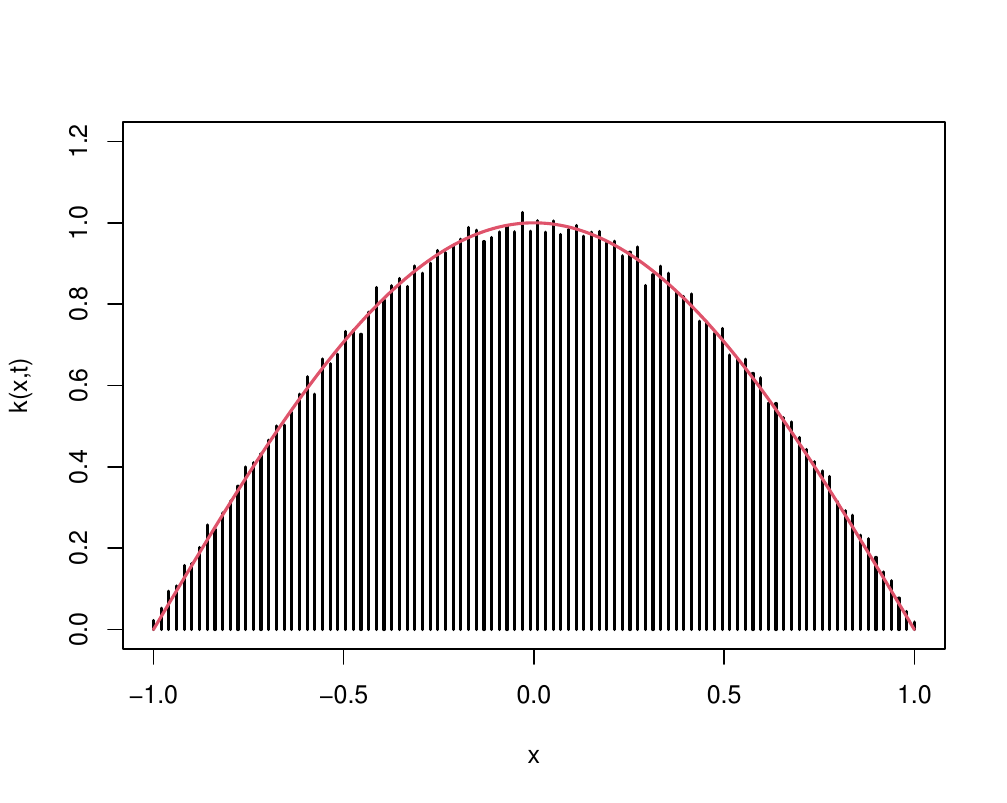}
\caption{Effects of the(tamed)  killing rate for the interval problem  with killing and branching: relaxation with the branching rate   ${\cal{V}}(x)= - \pi^2/8$ in terms of alive   trajectories counting.  In both  rows, time labels  appear  in the increasing order from left to right. }
\end{center}
\end{figure}
Note that scales  along the vertical axis, except for the first panel  are the same.   The  envelope (continuous curve) has an  exact  analytic form $k(x,t)$,  as  given by Eq. (35).    Compare e.g. the right panel of Fig. 2.
\begin{figure}[h]
\begin{center}
\centering
\includegraphics [width=0.4\columnwidth] {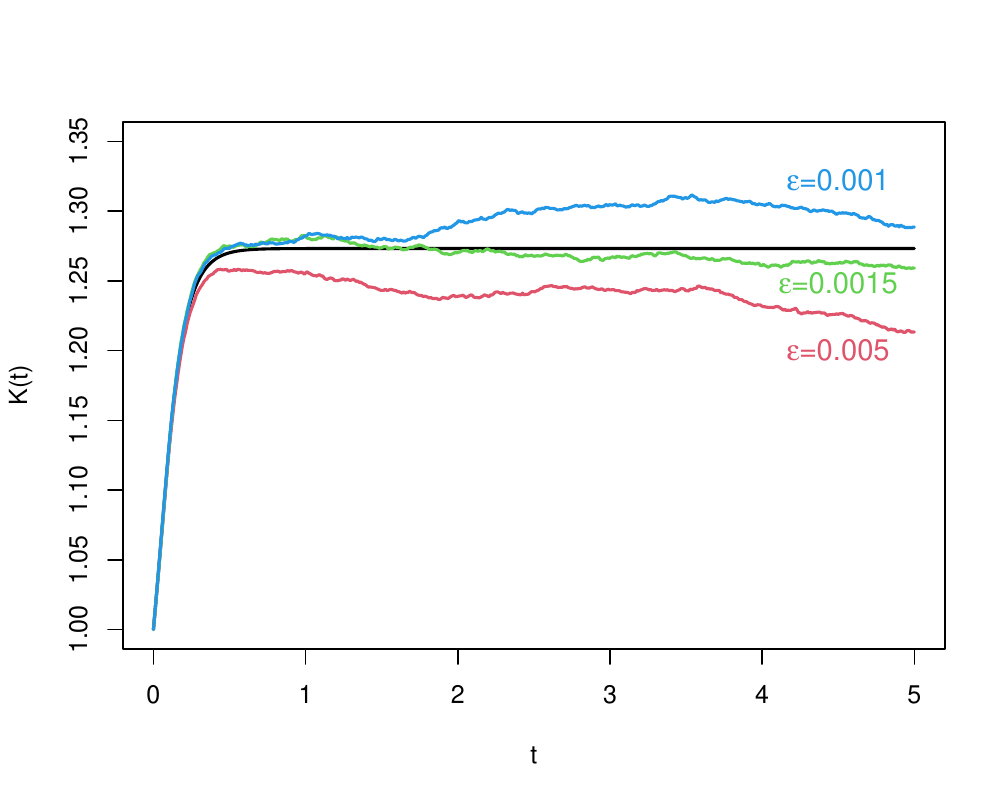}
\includegraphics [width=0.4\columnwidth] {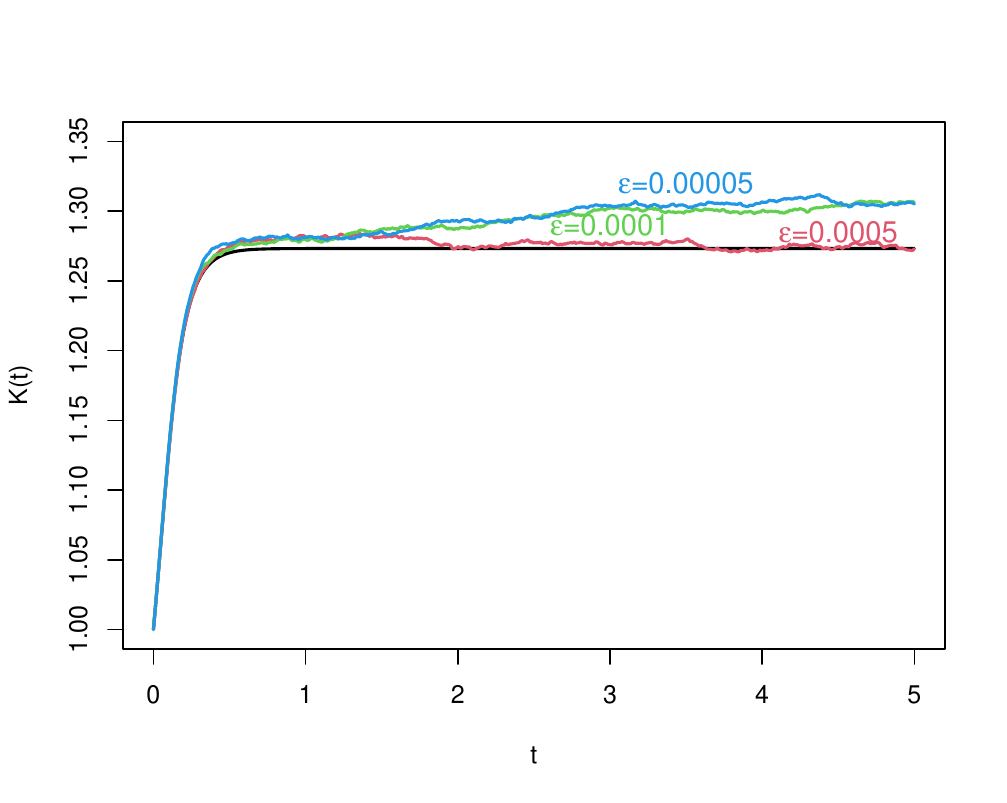}
\caption{Interval with absorbing ends. Branching  relaxation in terms of   $K(t)$, Eq. (38),  black curve,  set against numerically retrieved data  for $N(t)/N(0)  \approx K(t)$, with the predefined choice of  $\delta t $.   Left panel:  $\delta t= 0.0001$. The best  fit corresponds to the non-cloning treshold $\epsilon = 0.0015$  (green curve). We display  results of the enlargement of  this fit  to the  value $0.005$ and  its minimization to  $0.001$.  Right panel:  $\delta t= 0.00001$. By decreasing $10$ times the  previous  $\delta t$  we  pass from  $\epsilon = 0.0015$ to $\epsilon =0.0005$ (red).  We demonstrate that  the best  fit  cannot be arbitrarily minimized with $\delta t$ predefined and fixed.   We display deviations of   $N(t)/N(0)$  from   $K(t)$,  for test tresholds $ 0.0001$  and $0.00005$ . }
\end{center}
\end{figure}

We note that $K(t) \approx  N(t)/N(0) $ asymptotically approaches  $4/\pi \approx 1.27 $, and accordingly    $ N(t) \rightarrow  10^5 (4/\pi ) \approx 127 000$, which is a definite branching surplus effect, if compared with $N(0)= 10 ^5$.

  In case  of the interval with absorbing ends, all simulation runs are accomplished in the open interval $(-1,1)$, where there is no killing, while branching is allowed  at a constant rate.  All trajectories passing (reaching)  the endpoints $\pm 1$ of the interval are  abruptly killed.  At the boundaries, we must overcome the singularity of the model  (QM by provenance) infinite well  potential and the killing/branching  "abruptness" issue for the diffusion problem.

  We have found necessary to allow  trajectory cloning, which is actually  admitted   in $(-1,1)$ with a constant  probability $(\pi^2/8) \delta t$,  in the slightly reduced  domain   $[-1+\epsilon, 1-\epsilon]$, instead of $(-1,1)$ proper.  Told otherwise,  we do not admit  cloning in the $\epsilon $ vicinity of the endpoints $\pm 1$.

    To  have some control over the obvious loss of some branching trajectories  in the overall trajectory statistics,  we  test a number of the  parameter $\epsilon $   adjustments, to get the best fit to the analytical data of Section II.
   We have found that the  value  of $\epsilon $ can  be significantly  lowered  only  in strict  correlation with the improving finesse of the time scale coarse-graining ($\delta t = T/n$ with  $n$ increasing).

We have  comparatively displayed in Figs. 7  and 8   the analytic and computer-retrieved  trajectory counting data for the interval with absorbing ends, with  the cloning rate ${\cal{V}}= - \pi ^2 /8$.  In Fig. 8 the analytic curve $K(t)$ (black) satisfactorily  agrees with the computer-assisted data, if the best fit for $\epsilon$ is found, given the value of $\delta t$.  With the  choice of  $\delta t = 10^{-5}$  the best fit is  $\epsilon = 0.0005 $, c.f. the right panel of Fig. 8.

 One can see in  both panels of  Fig 8, how slight changes of $\epsilon $, which enlarge  or diminish the non-cloning area (and thus affect the reproduction  of trajectories),  lead to quite serious deviations from the analytic prediction $K(t)$. Specifically, we cannot freely minimize $\epsilon $, with  $\delta t$ predefined and fixed.

In the  left   panel of  Fig 8,   the optimal fit of the non-cloning treshold $\epsilon = 0.0015$  has been set  for  $\delta t =0.0001$. In the  right panel of Fig. 8,  where $\delta t=0.00001$   allows to minimize  the optimal treshold value down to $\epsilon = 0.0005$.
We expect that  with  further lowering of $\delta t$, the treshold value of $\epsilon $  would go down as well, thus increasing the accuract with which $N(t)/N(0)$ approximates the analytic outcome $K(t)$, (38).

 We have verified  what  might   happen,  if the non-cloning area is narrowed below the  optimal  value $\epsilon  = 0.0005$, while preserving  $\delta t = 10^{-5}$.
By  selecting   $\epsilon = 0.0001$ and $\epsilon =0.00005$, we obtain a definite  increase of the number of finally  recorded   trajectories. This  trajectories  surplus  is clearly identfiable beginning from  $t \approx  0.5$, once  compared with the  analytically known  behavior   of $K(t)$. We find convincingly  confirmed,  that   the  optimal  $N(t)/N(0) \approx K(t) $ curve   corresponds to $\epsilon \approx  0,0005$, albeit some (hyper)fine tuning  might  still be possible.

\section{Outlook.}

The above discusssion, both analytic and computer-assisted, of the branching/killing  path-wise  scenario for a continuous  passsage   from $y$ at $t=0$ to $x$ at $t>0$,  (clearly  realizable  along the  admissible, with  dead ends bypassed,   branching  trajectory)  has been motivated  out of curiosity.  Namely, while the renormalised Hamiltonians, \cite{glimm,faris},  with a conspicuous  shift-down  of the  potential function  on the energy scale,  have been  often employed in the literature, due to the  inflicted  "tamed killing effects" \cite{zaba} in  the  behavior of kernel functions $\exp(-tH)(y,x)$,  somewhat unexpectedly  nothing has been said about the intrinsic stochastic background of that behavior.

Relaxing (conditioned) diffusion processes  are described in terms of  Fokker-Planck  transition probability densities, whose main building blocks acually are the associated  Feynman-Kac path integral kernels. These kernels share the relaxation property as well, and that  quite often  due to the presence in the Feynman-Kac exponent,  of {\it killing rate (potential) functions with subtraction}, \cite{faris,zaba},  check   e.g.  \cite{stef} for an alternative viewpoint.

Here,  if  a nonegative  $V(x)$ would imply a standard killing scenario, \cite{faris,klauder}, by performing  a potentially trivial subtraction we  induce  quite nontrivial (branching picture) consequences. The main point here is that $H= -(1/2)\Delta + [V(x)] \rightarrow H= -(1/2)\Delta + [V(x) - E_0]$  places  $H$  in  the considered before (much broader)   family $H= -(1/2)\Delta + {\cal{V}}$ of Hamiltonian operators with the vanishing (equal $0$),  isolated  ground state eigenvalue, c.f. \cite{stef}.

We have quite intentionally skipped routine  introductory  phrases,  with a message about an  overall  relevance and  unquestionably  broad significance of   (i) the killing and mass creation in the Brownian motion, \cite{ito}-\cite{nagasawa1},  (ii)  branching stochastic processes, (iii) quasistationary distributions,  (iv) links with the Feynman-Kac formalism,  all that  within  physics or beyond physics (social sciences, biology etc.), since these can be found in  Refs. \cite{monthus}-\cite{faris} and furthermore in   Refs. \cite{huillet}-\cite{collet}.

The trajectory picture we have described in the present paper, to some extent may be interpreted in terms of the  metaphor, \cite{borges}, concerning ways allowing  to reach a predefined destiny  (here a terminal point $x$ at $t$), from a predefined beginning (starting point $y$ at $t=0$),  along a continuous  path, with branching/killing  events to be  observed  on the way.  We note that  a continuity  property  of the ultimate (uninterrupted) path, is nonetheless preserved.  This is not  to be mixed with the "the garden of forking paths fallacy" in statistics, \cite{gelman}.

\end{document}